\documentclass[12pt]{article}
\usepackage{a4wide,amssymb,cite}
\pdfoutput=1

\usepackage{a4wide,amssymb,graphicx}
\usepackage{epsfig}
\usepackage[usenames,dvipsnames]{color}
\usepackage{slashed}
\usepackage{amssymb,cite,graphicx}
\usepackage{slashed}
\usepackage{amsmath,bm,bbm}
\usepackage{amsfonts}
\usepackage[titletoc,title]{appendix}
\usepackage[small]{caption}
\usepackage[margin=1in]{geometry}
\usepackage[multiple]{footmisc}
\usepackage{mathtools}
\usepackage{slashed}
\usepackage{tabularx}
\usepackage[nottoc]{tocbibind}
\usepackage{xcolor}

\newcommand{\be}{\begin{equation}}
\newcommand{\ee}{\end{equation}}
\newcommand{\bea}{\begin{eqnarray}}
\newcommand{\eea}{\end{eqnarray}}

\def\circa#1{\,\raise.3ex\hbox{$#1$\kern-.75em\lower1ex\hbox{$\sim$}}\,}

\begin{document}

\begin{titlepage}
%
%


%

\begin{centering}
\vspace{1cm}
{\Large {\bf Inflation models with Peccei-Quinn symmetry \vspace{0.15cm} \\  and axion kinetic misalignment}} \\

\vspace{1.5cm}

\begin{centering}
{\bf Hyun Min Lee$^{1\dagger}$, Adriana G. Menkara$^{1,2\sharp}$,  Myeong-Jung Seong$^{1\ddagger}$, \vspace{0.15cm} \\  and Jun-Ho Song$^{1\star}$ }
\end{centering}
\\
\vspace{.5cm}

{\it $^1$Department of Physics, Chung-Ang University, Seoul 06974, Korea.} \\ \vspace{.2cm}
{\it $^2$Deutsches Elektronen-Synchrotron DESY, Notkestrasse 85, 22607 Hamburg, Germany.}

\vspace{.5cm}


\end{centering}
\vspace{1.5cm}

\begin{abstract}
\noindent
We propose a consistent framework with the $U(1)$ Peccei-Quinn (PQ) symmetry for obtaining the initial condition for axion kinetic misalignment from inflation. We introduce a PQ complex scalar field and an extra Higgs doublet, which are conformally coupled to gravity, and three right-handed neutrinos for the seesaw mechanism. In the DFSZ type scenarios for the axion, we obtain the PQ anomalies from the Standard Model quarks carrying nonzero PQ charges in some of two Higgs doublet models, solving the strong CP problem by the QCD potential for the axion. Assuming that the PQ symmetry is explicitly violated in the scalar potential by quantum gravity effects, we show that a sufficiently large initial axion velocity can be obtained before the QCD phase transition while avoiding the axion quality problem. 
As inflation is driven by the radial distance from the origin in the space of scalar fields close to the pole of the kinetic terms in the Einstein frame, we obtain successful inflationary predictions and set the initial axion velocity at the end of inflation. Focusing on the pure PQ inflation with a small running quartic coupling for the PQ field, we discuss the post-inflationary dynamics for the inflaton and the axion. As a result, we show that  a sufficiently high reheating temperature can be obtained dominantly from the Higgs-portal couplings to the PQ field, while being consistent with axion kinetic misalignment, the stability for the Higgs fields during inflation and the non-restoration of the PQ symmetry after reheating.

\end{abstract}

\vspace{2cm}

\begin{flushleft} 
$^\dagger$Email: hminlee@cau.ac.kr \\
${}^{\sharp}$Email: adriana.menkara@desy.de \\
$^\ddagger$Email: tjdaudwnd@gmail.com  \\
$^\star$Email: thdwnsgh1003@naver.com
\end{flushleft}

\end{titlepage}

\section{Introduction}

The QCD axion is a pseudo-Goldstone boson of the $U(1)$ Peccei-Quinn (PQ) symmetry and it solves the strong CP problem in the Standard Model (SM) by the dynamical relaxation mechanism during the cosmological evolution after the QCD phase transition \cite{kim-review}. It is also a candidate for decaying cold dark matter \cite{misalign}, with feeble interactions to gluons and photons and potentially other SM particles \cite{ksvz,dfsz}. The relic abundance for the axion is determined by the misalignment mechanism \cite{misalign} where the axion is assumed to be displaced from the origin of the axion potential during or after inflation and becomes a coherently oscillating cold dark matter after the QCD phase transition. 

Recently, the possibility that the axion has a nonzero initial velocity before the QCD phase transition has drawn much attention \cite{Co:2019jts}. In this case, even after the QCD phase transition, the axion is not trapped into one of the minima of the periodic potential until its kinetic energy is comparable to the axion potential energy. Thus, the axion oscillation is delayed until a late time after the standard oscillation time of the axion without a initial velocity, so the dilution of the axion energy density due to the Hubble expansion is milder, leading to the correct relic density even for a smaller initial energy density of the axion or a smaller axion decay constant \cite{Co:2019jts}. In order to achieve a nonzero initial velocity for the axion, however, an explicit violation of the PQ symmetry is required, while the axion quality is maintained to solve the strong CP problem. Thus, it is necessary to make the overall coefficient of the PQ violating potential dynamical, so there must be a period in the early universe where the radial mode of the PQ complex scalar field is much larger than the one in the vacuum and it decays away \cite{Co:2019jts,PQpole,Eroncel:2024rpe}. 

In this article, we aim to achieve a nonzero initial axion velocity for the axion kinetic misalignment in a consistent model with the $U(1)$ PQ symmetry for inflation, reheating and all the way to the axion oscillation. We extend our previous results of the PQ inflation in KSVZ type axion models \cite{PQpole} to the case in DFSZ type axion models where one more Higgs doublet and the PQ complex scalar field are added beyond the SM \cite{dfsz}, instead of an extra heavy vector-like colored quark. We also introduce the interactions for three right-handed (RH) neutrinos being consistent with the PQ symmetry and obtain the masses for the RH neutrinos due to the spontaneous breaking of the PQ symmetry. Thus, neutrino masses can be generated in the seesaw mechanism with a small mixing between the active neutrinos and the heavy RH neutrinos.

We assume that the PQ field and two Higgs doublets are coupled conformally to gravity and the scalar potential is composed of PQ invariant renormalizable terms and PQ violating higher dimensional terms. In this setup, we derive the effective Lagrangians for the inflaton in the cases for both the Higgs-PQ mixed and pure PQ inflations.  The inflaton is identified as the radial distance from the origin in the field place of scalar fields, driving a slow-roll inflation close to the pole of the kinetic terms \cite{Higgspole}.  We also consider the stabilization of all the non-inflaton scalar fields during inflation and identify the initial axion velocity at the end of inflation. 

Focusing on the pure PQ inflation where the correct inflation scale is achieved from a small running quartic coupling of the PQ field, we analyze the inflationary predictions, the post-inflationary dynamics of the inflaton and the perturbative reheating, and determine the relic density from the axion kinetic misalignment.  We show how the stabilization conditions for the Higgs fields and the non-restoration of the PQ symmetry after reheating can constrain the parameter space for the axion kinetic misalignment.

The paper is organized as follows.
We begin with the setup for DFSZ type axion models and present two Higgs doublet models based on $Z_2$ parities and anomaly coefficients for the PQ symmetry in each model. Then, we discuss the vacuum structure of the model from the scalar potential for charge-neutral scalar fields and discuss the axion quality problem with general PQ violating potentials. Next, the detailed discussion on the effective theory for the inflaton with the Higgs-PQ mixed field or the pure PQ field is presented and the stabilization of the non-inflaton scalar fields is shown. Afterwards, we discuss the slow-roll inflation and the initial axion velocity in the case of the pure PQ inflation. We also consider the reheating procedure and determine the amount of dark radiation from the axions produced during reheating.  From the post-inflationary evolution of the axion velocity, we also get the relic density from the axion kinetic misalignment after the QCD phase transition. Finally, conclusions are drawn.

\section{The setup}

We present the model setup for the PQ inflation in the DFSZ axion model where there are a complex PQ scalar field and an extra Higgs doublet beyond the SM. We identify four types of two Higgs doublet models depending on $Z_2$ parities and include three right-handed neutrinos for getting neutrino masses by the seesaw mechanism, being consistent with the PQ symmetry. We also show the axion couplings to gluons and photon in each of two Higgs doublet models.

We first consider the bosonic part of the Lagrangian in the Jordan frame  in the DFSZ axion model as
\bea
\frac{{\cal L}_J}{\sqrt{-g_J}} = -\frac{1}{2}M^2_P\, \Omega R(g_J) + \mathcal{L}_{{\rm kin},J} -\Omega^2 V_E \label{LJ}
\eea
where $\Omega$ is the frame function of $H_1, H_2$ and $\Phi$, the kinetic terms are given by
\bea
 \mathcal{L}_{{\rm kin},J}  = |D_\mu H_1|^2  + |D_\mu H_2|^2 +  |\partial_\mu \Phi|^2
\eea
and $V_E$ is the Einstein frame potential, composed of PQ invariant and PQ violating terms as $V_E=V_{\rm PQ}+V_{\rm PQV}$.

Taking the conformal couplings of the scalar fields to gravity by
\bea
\Omega = 1 - \frac{1}{3M^2_P}|H_1|^2 -  \frac{1}{3M^2_P}|H_2|^2 -  \frac{1}{3M^2_P}|\Phi|^2, 
\eea
we obtain the  Einstein frame Lagrangian as
\bea  \label{LEin}
\frac{\mathcal{L}_{E}}{\sqrt{-g_E}} = -\frac{1}{2}M^2_P R + \frac{1}{\Omega}\left(|D_\mu H_1|^2 + |D_\mu H_2|^2 + |\partial_\mu \Phi|^2\right)  + \frac{3}{4}M^2_P \frac{\left(\partial_\mu \Omega\right)^2}{\Omega^2} -V_E.
\eea

\subsection{Scalar potential}

We assume that the scalar fields transform under the PQ symmetry as 
\bea
\Phi &\rightarrow& e^{i q_\Phi \alpha} \Phi,\nonumber\\
H_1 &\rightarrow& e^{i q_1 \alpha} H_1,\nonumber\\
H_2 &\rightarrow& e^{i q_2 \alpha} H_2.\nonumber
\eea
Then, the PQ-invariant Einstein frame potential is given by
\bea
V_{\rm PQ} &=& \lambda_\Phi |\Phi|^4 + \lambda_1 |H_1|^4 + \lambda_2 |H_2|^4 + \lambda_3|H_1|^2|H_2|^2 + \lambda_4\left(H_1^\dagger H_2\right)\left(H_2^\dagger H_1\right) \nonumber \\
&&+\lambda_{1\Phi} |H_1|^2|\Phi|^2 + \lambda_{2\Phi}|H_2|^2|\Phi|^2 + \left(2^{p/2-1}\kappa_p H_1^\dagger H_2 \Phi^p + \text{h.c.} \right) \nonumber \\
 &&+ \mu_\Phi^2 |\Phi|^2 + m_1^2 |H_1|^2 + m_2^2|H_2|^2+V_0, \label{VE}
\eea
with $V_0$ being a cosmological constant. 
Then, if the PQ charges satisfy $pq_\Phi -q_1 + q_2 =0$, the $H_1^\dagger H_2 \Phi^p $ term is PQ invariant\footnote{For renormalizable interactions, we can take $p=1$ or $p=2$.}.
Since the PQ symmetry is broken dominantly by the VEV of the $\Phi$ field for the invisible axion, we need to choose $q_\Phi\neq 0$.
We introduce a cosmological constant $V_0$ in the Einstein frame to set the vacuum energy to zero.

The PQ symmetry can be broken explicitly due to quantum gravity, with the PQ violating terms in the potential, as follows,
\bea
V_{\rm PQV} = \sum_{n,l}\sum_{k=0}^{[l/2]}{\frac{c_{n,l,k}}{2M^{2n+l-4}}}\, (H^\dagger_1 H_2)^n |\Phi|^{2k} \Phi^{l-2k} +{\rm h.c.}, \label{PQVp}
\eea
for $n(q_2-q_1)+q_\Phi (l-2k)\neq 0$, which are constrained by the quality of the axion for solving the strong CP problem.
Henceforth, we set $M_P=1$ for convenience, but we recover $M_P$ whenever necessary.

\subsection{Two Higgs doublet models}

The most general Yukawa interactions will be of the form
\bea
\mathcal{L}_Y = y_{ij} \bar{f}_LH_1 f_R  + y'_{ij} \bar{f}_LH_2 f_R,
\eea
up to the replacement of $H_{1,2}$ with ${\tilde H}_{1,2}$ if $f_R$ is an up-type quark or an RH neutrino.
However, the general Yukawa couplings would lead to dangerous FCNCs. Thus, there are several phenomenologically viable possibilities for the flavor-independent assignments for $Z_2$ parities and PQ charges, summarized in Table~\ref{PQcharges}. 

\begin{table}[hbt!]  
    \begin{center}
  \scalebox{0.9}{
   \begin{tabular}{c|cccccccc}  \hline\hline
    $Z_2$ & $\Phi$ & $H_1$ & $H_2$ & $q_L$ & $u_R$ & $d_R$ & $l_L$ & $e_R$   \\ \hline
     Type I & $+$ & $-$ & $+$ & $+$ & $+$ & $+$ & $+$ & $+$   \\\hline
     Type II & $+$ & $-$ & $+$ & $+$ & $+$ & $-$ & $+$ & $-$   \\\hline  
     Type X & $+$ & $-$ & $+$ & $+$ & $+$ & $+$ & $+$ & $-$  \\\hline
     Type Y & $+$ & $-$ & $+$ & $+$ & $+$ & $-$ & $+$ & $+$ \\    
      \hline\hline
   \end{tabular}}
   
    \begin{center}
  \scalebox{0.9}{
      \begin{tabular}{c|ccccccccc}
        \hline\hline 
        PQ & $\Phi$ & $H_1$ & $H_2$  & $q_L$ & $u_{R}$  &  $d_{R}$ & $l_{L}$ & $e_{R}$  \\\hline
        Type I & $q_\Phi$ &  $q_1$ & $q_1-p q_\Phi$ &  $0 $ & $q_1-pq_\Phi$ & $-q_1+pq_\Phi$ & $0$ & $0$  \\\hline
              Type II & $q_\Phi$ & $q_1$ & $q_1-pq_\Phi$ & $0$ & $q_1-pq_\Phi$ & $-q_1$ & $0$ & $-q_1$   \\ \hline
              Type X & $q_\Phi$ & $q_1$ & $q_1-pq_\Phi$ & $0$ & $q_1-pq_\Phi$ & $-q_1+pq_\Phi$ & $0$ & $-q_1$  \\\hline
              Type Y & $q_\Phi$ & $q_1$ & $q_1-pq_\Phi$ & $0$ & $q_1-pq_\Phi$ & $-q_1$ & $0$ & $-q_1+pq_\Phi$  \\ 
        \hline\hline
      \end{tabular}} \\[4mm]  
    \end{center}

    \end{center}
      \caption{$Z_2$ parities and PQ charges  for two Higgs doublet models. \label{PQcharges}}
  \end{table}

Looking at Table \ref{PQcharges}, we see that for the Type I model, all quarks couple to just one of the Higgs fields, which we take to be $H_2$. 
In contrast, in the Type II model, $u_R$ couples to $H_2$ but $d_R$ and $e_R$ couple to $H_1$.
In the Type X, both $u_R$ and $d_R$ couple to $H_2$ but $e_R$ couples to $H_1$, and finally, in the Type Y (flipped), $u_R$ and $e_R$ couple to $H_2$ and $d_R$ couples to $H_1$. 

For instance, in the case of the Type II model in Table~\ref{PQcharges},  we obtain the PQ and $Z_2$ invariant Yukawa couplings,
\bea
\mathcal{L}_Y = - Y_u\overline{q}_L \tilde{H}_2u_R - Y_d \overline{q}_L H_1 d_R - Y_e \overline{l}_LH_1e_R.
\eea

We comment on the neutrino Yukawa couplings and the mass terms for the right-handed (RH) neutrinos, $N_R$. 
If the RH neutrinos are $Z_2$-even, we can introduce the neutrino Yukawa couplings for Dirac and Majorana neutrino masses, as follows,
\bea
\mathcal{L}^{(1)}_\nu = -Y_\nu \overline{l}_L \tilde{H}_2 N_R- \frac{1}{2} y_N \overline{N^c_R} \Phi N_R.
\eea
Then, we need to impose $q_{N_R}=q_1-pq_\Phi=-\frac{1}{2}q_\Phi$, so $q_\Phi=\frac{2}{2p-1} q_1$ and $q_{N_R}=-\frac{q_1}{2p-1}$.
On the other hand, if the RH neutrinos carry $Z_2$-odd parities, we need to take alternative mass terms for neutrinos, as follows,
\bea
\mathcal{L}^{(2)}_\nu = -Y_\nu \overline{l}_L \tilde{H}_1 N_R- \frac{1}{2} y_N \overline{N^c_R} \Phi N_R
\eea
In this case, we need to take the PQ charge for $N_R$ as $q_{N_R}=q_1=-\frac{1}{2}q_\Phi$ instead of $q_{N_R}=q_1-pq_\Phi$ in Table \ref{PQcharges}, so $q_\Phi=-2q_1$ and $q_{N_R}=q_1$.

\subsection{Axion couplings to gluons and photon}

The color and electromagnetic anomalies of the axial current associated with  the axion are given in the following Lagrangian,
\bea
S_{\rm anom}=\frac{k_G}{8\pi^2}\int  \theta\, {\rm tr}(G\wedge G)+\frac{k_F}{8\pi^2}\int \theta\,  F\wedge F,
\eea
with $\theta=a/f_a$. Then, we obtain the effective axion-photon coupling below the PQ symmetry breaking scale, as follows,
\bea
{\cal L}_{\rm photon} =-\frac{1}{4} g_{a\gamma\gamma} \, a\, F_{\mu\nu} {\tilde F}^{\mu\nu}
\eea
with
\bea
g_{a\gamma\gamma} =\frac{\alpha}{2\pi f_a/\xi} \bigg(\frac{E}{N}-1.92\bigg).
\eea
Here, $\xi=k_G$ and $E/N=2k_F/k_G$ from $N=\frac{1}{2}k_G$ and $E=k_F$, and the domain wall number is given by $N_{\rm DW}=\xi=k_G$ \cite{kim-review}.  The values for $k_G, k_F$ in DFSZ-like models \cite{dfsz} are shown in Table~\ref{Anomalies}. Here, we note that $E/N=\frac{8}{3}$ for Type II, which is consistent with unification. For comparison, in KSVZ model with a neutral vector-like colored triplet \cite{ksvz}, the anomaly coefficients are given by $k_G=1$ and $k_F=0$, so $E/N=0$.

\begin{table}[hbt!]  
    \begin{center}
  \scalebox{0.9}{
   \begin{tabular}{c|ccc}  \hline\hline
     & $k_G$ & $k_F$  & $E/N$ \\ \hline
     Type I & $0$ & $3(pq_\Phi-q_1)$ & $-$  \\\hline
     Type II & $3pq_\Phi$ & $4p q_\Phi$  & $\frac{8}{3}$ \\\hline  
     Type X & $0$ & $3pq_\Phi$  & $-$ \\\hline
     Type Y & $3pq_\Phi$ & $pq_\Phi$  & $\frac{2}{3}$ \\    
      \hline\hline
   \end{tabular}}
    \end{center}
      \caption{PQ anomalies and axion-photon couplings in two Higgs doublet models. \label{Anomalies}}
  \end{table}

It was recently pointed out that if there is no fractionally charged color-singlet particle, the SM gauge groups are globally identified by $[SU(3)_C\times SU(2)_L\times U(1)_Y]/Z_6$, which corresponds to $[SU(3)_C\times U(1)_{\rm em}]/Z_3$ after electroweak symmetry breaking. In this case, the anomaly coefficients, $k_G, k_F$, are quantized by $k_G\in Z$ and $\frac{2}{3}k_G+k_F\in Z$ \cite{quantization,quant2}. The first quantization condition means $3pq_\Phi\in Z$ for both Type II and Y. On the other hand, since $\frac{2}{3}k_G+k_F=6pq_\Phi, 3pq_\Phi$ for Type II and Y, respectively, we find that an integer, $k_G=3pq_\Phi$, is sufficient for the second quantization condition. If $q_\Phi$ is an integer, which is a sufficient condition for the first quantization condition, the domain wall number is a multiple of three or six from $N_{\rm DW}=k_G=3pq_\Phi$ for $p=1$ or $p=2$. So, if the PQ symmetry is spontaneously broken after reheating, there is a notorious domain wall problem in this case. However, if we tolerate the quantization condition to an integer $k_G$, we can take $q_\Phi$ as a multiple of $1/(3p)$, avoiding the domain wall problem in the standard scenarios \footnote{Nonetheless, the domain wall problem reappears due to non-invertible domain walls \cite{quant2}.}. We note that  $E/N$ is the same, independent of $p$ in the PQ-invariant potential, $H^\dagger_1 H_2 \Phi^p$.

\section{PQ symmetry breaking and axion quality problem}

We consider the Lagrangian for charge-neutral scalars in the Einstein frame in order to discuss the spontaneous breaking of electroweak symmetry and PQ symmetry. We show the conditions for the stable electroweak vacuum in the model and discuss the axion quality problem in the presence of higher dimensional PQ violating terms in the potential.

\subsection{Einstein frame Lagrangian for neutral scalars}

We parametrize the charge-neutral components of Higgs and the PQ fields in the following,
\bea
H_1 = \frac{1}{\sqrt{2}} \begin{pmatrix} 0 \\ h_1e^{i\eta_1} \end{pmatrix},  \quad
H_2 = \frac{1}{\sqrt{2}} \begin{pmatrix} 0 \\ h_2 e^{i\eta_2} \end{pmatrix}, \quad
\Phi = \frac{1}{\sqrt{2}}\,\rho \, e^{i\theta}.  \label{polarrep}
\eea
Then, the kinetic terms of the Einstein-frame Lagrangian in eq.~(\ref{LEin}) become
\bea 
\frac{\mathcal{L}_{\mathrm{kin},E}}{\sqrt{-g_E}} &=&  -\frac{1}{2}R + \frac{ \left(\partial_\mu h_1\right)^2 + \left(\partial_\mu h_2\right)^2 + h_1^2\left(\partial_\mu \eta_1\right)^2 + h_2^2\left(\partial_\mu \eta_2\right)^2 + \left(\partial_\mu \rho\right)^2 + \rho^2\left(\partial_\mu \theta\right)^2}{2(1 - \frac{1}{6}h_1^2 - \frac{1}{6} h_2^2 -\frac{1}{6}\rho^2)} \nonumber\\
&&\quad+ \frac{3}{4} \frac{\Big(\partial_\mu\left(\frac{1}{6}h_1^2 + \frac{1}{6}h_2^2 + \frac{1}{6}\rho^2\right)\Big)^2}{( 1- \frac{1}{6}h_1^2 - \frac{1}{6} h_2^2 -\frac{1}{6}\rho^2)^2}. \label{Ein2}
\eea
Moreover, we can rewrite the PQ-invariant Einstein-frame potential for neutral scalars from eq.~(\ref{VE}) as
\bea
V_{\rm PQ} &=&V_0+\frac{1}{2} m^2_1 h^2_1 +\frac{1}{2}m^2_2 h^2_2  +\frac{1}{4} \lambda_1 h^4_1 + \frac{1}{4} \lambda_2 h^4_2 + \frac{1}{4} (\lambda_3+\lambda_4) h^2_1 h^2_2 \nonumber \\
&&+\frac{1}{2}\mu^2_\Phi \rho^2+ \frac{1}{4} \lambda_\Phi \rho^4 +\frac{1}{4} \lambda_{1\Phi} h^2_1 \rho^2  +\frac{1}{4} \lambda_{2\Phi} h^2_2 \rho^2 +\frac{1}{2}   \kappa_p h_1 h_2 \rho^p \cos(\eta_2-\eta_1+p\theta),
\label{PQ}
\eea
and the PQ violating Einstein-frame potential in eq.~(\ref{PQVp}) as
\bea
V_{\rm PQV} =\sum_{n,l}\sum_{k=0}^{[l/2]} \frac{|c_{n,l,k}|}{2^{n+l/2}}\, h^n_1 h^n_2 \rho^l \cos\Big[n(\eta_2-\eta_1+p\theta)+(l-2n-2k)\theta+A_{n,l,k}\Big],
\label{PQV}
\eea
with $A_{n,l,k}$ being constant phase shifts.

\subsection{Vacuum structure}

We determine the vacuum structure of the model for breaking the PQ and electroweak symmetries simultaneously. 

As the VEVs of the PQ and Higgs fields are much smaller than the Planck scale, we can approximate the kinetic terms in the Einstein-frame in eq.~(\ref{Ein2}) to be almost of canonical form. In order to determine the vacuum, we consider the minimization conditions for the PQ-invariant potential assuming that the PQ violating potential is small enough for axion quality problem. Minimizing one of the angular modes by $\eta_2-\eta_1+p\theta$ from the cosine potential in eq.~(\ref{PQ}), we get the PQ-invariant potential in the following form,
\bea
V_{\rm PQ}= &=&V_0+\frac{1}{2} m^2_1 h^2_1 +\frac{1}{2}m^2_2 h^2_2  +\frac{1}{4} \lambda_1 h^4_1 + \frac{1}{4} \lambda_2 h^4_2 + \frac{1}{4} (\lambda_3+\lambda_4) h^2_1 h^2_2 \nonumber \\
&&+\frac{1}{2}\mu^2_\Phi \rho^2+ \frac{1}{4} \lambda_\Phi \rho^4 +\frac{1}{4} \lambda_{1\Phi} h^2_1 \rho^2  +\frac{1}{4} \lambda_{2\Phi} h^2_2 \rho^2 -\frac{1}{2}    |\kappa_p| h_1 h_2 \rho^p.
\eea
Then, the minimization conditions for the PQ-invariant potential are given by
\bea
&&\mu^2_\Phi \rho +\lambda_\Phi \rho^3 + \frac{1}{2} \lambda_{1\Phi} h^2_1 \rho+\frac{1}{2} \lambda_{2\Phi} h^2_2 \rho- \frac{1}{2}p|\kappa_p | h_1 h_2 \rho^{p-1}=0, \label{min1} \\
&&m^2_1 h_1 +\lambda_1 h^3_1 +  \frac{1}{2} (\lambda_3+\lambda_4) h_1 h^2_2+\frac{1}{2} \lambda_{1\Phi} h_1 \rho^2-\frac{1}{2}    |\kappa_p|  h_2 \rho^p=0, \label{min2} \\
&&m^2_2 h_2 +\lambda_2 h^3_2 +  \frac{1}{2} (\lambda_3+\lambda_4) h^2_1 h_2+\frac{1}{2} \lambda_{2\Phi} h_2 \rho^2-\frac{1}{2}    |\kappa_p|  h_1 \rho^p=0. \label{min3}
\eea
We denote the VEVs by $\langle h_1\rangle=v_1=v\cos\beta$ and $\langle h_2\rangle=v_2=v\sin\beta$, with $v^2=v^2_1+v^2_2$, and  $\langle\rho\rangle\equiv v_\Phi$. Then, we get a nonzero VEV for the PQ field from eq.~(\ref{min1}) as
\bea
v_\Phi=\sqrt{\frac{1}{\lambda_\Phi} \Big(-\mu^2_\Phi -\frac{1}{2}\lambda_{1\Phi} v^2\cos^2\beta -\frac{1}{2}\lambda_{2\Phi} v^2 \sin^2\beta+ \frac{1}{2}|\kappa_2| v^2\sin2\beta  \Big)}, \label{PQvev}
\eea
for $p=2$, and $v_\Phi$ is a solution to the cubic equation in eq.~(\ref{min1}) for $p=1$. 
We can also rewrite eqs.~(\ref{min2}) and (\ref{min3}) into the equations determining the electroweak scale $v$ and $\sin2\beta$ in terms of the effective Higgs mass parameters and quartic couplings, as follows,
\bea
\frac{v^2}{2} = \frac{(1-\cos2\beta)m^2_{2,{\rm eff}}-(1+\cos2\beta) m^2_{1,{\rm eff}}}{(1+\cos2\beta)^2\lambda_1-(1-\cos2\beta)^2\lambda_2},
\label{weakscale}
\eea
and
\bea
\sin2\beta \bigg(m^2_{1,{\rm eff}}+m^2_{2,{\rm eff}}+\frac{1}{2}(\lambda_1+\lambda_2+\lambda_3+\lambda_4) v^2+\frac{1}{2}(\lambda_1-\lambda_2) v^2 \cos2\beta \bigg) = |\kappa_p| (v_\Phi)^p. \label{vevratio}
\eea
with $m^2_{1,{\rm eff}}\equiv m^2_1+\frac{1}{2}\lambda_{1\Phi} v^2_\Phi$ and $m^2_{2,{\rm eff}}\equiv m^2_2+\frac{1}{2}\lambda_{2\Phi} v^2_\Phi$ being the effective Higgs mass parameters. We note that the electroweak scale is controlled by $m^2_{i,{\rm eff}}$ and $\lambda_i$ ($i=1,2$) for a given $\sin2\beta$.  For $\kappa_p=0$, we get $\sin2\beta=0$, so only the Type I 2HDM would be a viable option for getting all the SM fermion masses. But, in this case, there is no QCD anomaly for the PQ symmetry, so there is no axion solution to the strong CP problem. Thus, we need a nonzero $\kappa_p$ for realizing the consistent electroweak symmetry breaking and the QCD axion, in particular, in Type II and Y 2HDMs where the QCD anomalies for the PQ symmetry are nonzero.  

We remark that the linear combination of $h_1$ and $h_2$ has a tachyonic mass near the origin for electroweak scale. Namely, for  ${\cal M}^2$ being the squared mass matrix for $h_1$ and $h_2$, we need ${\rm det} {\cal M}^2<0$, resulting in the following necessary condition,
\bea
m^2_{1,{\rm eff}} m^2_{2,{\rm eff}}< \frac{1}{4} |\kappa_p|^2 (v_\Phi)^{2p}. \label{cond1}
\eea
Otherwise, $h_1=h_2=0$ would be a stable minimum of the potential and electroweak symmetry breaking would not occur.
The above condition is automatically satisfied if the signs of $m^2_{1,{\rm eff}}$ or $m^2_{2,{\rm eff}}$ are opposite. Otherwise, the effective mass parameters are bounded.

We also note that eq.~(\ref{vevratio}) with $|\sin2\beta|< 1$ leads to the upper bound on $|\kappa_p|$, as follows,
\bea
 |\kappa_p| (v_\Phi)^p<m^2_{1,{\rm eff}}+m^2_{2,{\rm eff}}+\frac{1}{2}(\lambda_1+\lambda_2+\lambda_3+\lambda_4) v^2+\frac{1}{2}(\lambda_1-\lambda_2) v^2 \cos2\beta. \label{cond2}
\eea
Thus, the effective mass parameters for the Higgs doublets, $m^2_{1,{\rm eff}}$, $m^2_{2,{\rm eff}}$ and $ |\kappa_p| (v_\Phi)^p$, are constrained by eqs.~(\ref{cond1}) and (\ref{cond2}), apart from the condition for the electroweak scale in eq.~(\ref{weakscale}).

We comment on the upper bounds on the couplings between the Higgs doublets and the PQ field from electroweak symmetry breaking. First, the effective mass parameters, $m^2_{1,{\rm eff}}, m^2_{2,{\rm eff}}$, should not be far from of the weak scale, to get the electroweak symmetry scale in  eq.~(\ref{weakscale}) without fine-tuning parameters. Then, the mixing quartic couplings, $\lambda_{1\Phi}, \lambda_{2\Phi}$, contributing to the the effective mass parameters, should be sufficiently small for $v \ll v_\Phi$. For instance,  for $v_\Phi\sim 10^8\,{\rm GeV}$ and $|m^2_{1,{\rm eff}}|, |m^2_{2,{\rm eff}}|\lesssim (1\,{\rm TeV})^2$, we need $|\lambda_{1\Phi}|, |\lambda_{2\Phi}|\lesssim 10^{-10}$. Moreover, eq.~(\ref{vevratio}) is satisfied if $ |\kappa_p| (v_\Phi)^p\lesssim |m^2_{1,{\rm eff}}|, |m^2_{2,{\rm eff}}|\lesssim (1\,{\rm TeV})^2$, so we need $|\kappa_2|\lesssim 10^{-10}$ for $p=2$ or $|\kappa_1|\lesssim 10^{-2}\,{\rm GeV}$ for $p=1$.  
If there is fine-tuning between larger values of $m^2_{1,{\rm eff}}$ and $m^2_{2,{\rm eff}}$ in eq.~(\ref{weakscale}), we can allow for larger values of  $|\lambda_{1\Phi}|, |\lambda_{2\Phi}|$ and $ |\kappa_p| (v_\Phi)^p$, as far as eq.~(\ref{vevratio}) is satisfied. In this work, we take  $|\lambda_{1\Phi}|, |\lambda_{2\Phi}|, \kappa_p$ as being the free parameters as far as the running quartic coupling $\lambda_\Phi$ is small enough for inflation. So, we need to take $|\lambda_{1\Phi}|, |\lambda_{2\Phi}|, |\kappa_2|\lesssim10^{-5}$ and eq.~(\ref{cond2}) leads to $|\kappa_1|\lesssim (m^2_{1,{\rm eff}}+m^2_{2,{\rm eff}})/v_\Phi\lesssim \frac{1}{2}(\lambda_{1\Phi} + \lambda_{2\Phi})v_\Phi\lesssim 10^3\,{\rm GeV}$ for $v_\Phi\sim 10^8\,{\rm GeV}$.

\subsection{Axion quality problem}

In order to discuss the axion quality problem in the presence of PQ violating terms in the potential, we expand the Higgs and PQ fields around their VEVs, as follows,
\bea
H_1 = \frac{1}{\sqrt{2}} \begin{pmatrix} 0 \\ (v_1+\rho_1)e^{ia_1/v_1} \end{pmatrix},  \quad
H_2 = \frac{1}{\sqrt{2}} \begin{pmatrix} 0 \\ (v_2+\rho_2) e^{ia_2/v_2} \end{pmatrix}, \quad
\Phi = \frac{1}{\sqrt{2}}(v_\Phi+s) \, e^{ia_\Phi/v_\Phi}.
\eea
Then, among the neutral Goldstone bosons,  $a_1, a_2, a_\Phi$, we can identify the would-be Goldstone boson $G$, eaten by the $Z$ boson, and the heavy pseudo-scalar $A$, and the remaining orthogonal pseudo-scalar ${\bar a}$, as follows,
\bea
G &=& \frac{1}{v} (v_1 a_1 + v_2 a_2), \label{Goldstone} \\
A&=& N \bigg(\frac{a_2}{v_2}-\frac{a_1}{v_1} +\frac{pa_\Phi}{v_\Phi} \bigg), \label{pseudoscalar} \\
{\bar a}&=&K \bigg(v_2 a_1-v_1 a_2 +\frac{v_\Phi v^2}{pv_1 v_2}a_\Phi\bigg),
\eea 
with  $v=\sqrt{v^2_1+v^2_2}$, $N=1/\sqrt{v^{-2}_1+v^{-2}_{2}+p^2 v^{-2}_\Phi}$ and $K=1/\sqrt{v^2_1+v^2_2+ v_\Phi^2 v^4/(p^2v^2_1 v^2_2)}$.

On the other hand, the QCD axion is identified from the PQ Noether current, as follows,
\bea
a &=&\frac{1}{f_a}\, \Big( q_1 v_1 a_1 + q_2 v_2 a_2 + q_\Phi v_\Phi a_\Phi  \Big) \label{QCDaxion}
\eea
where $f_a=\sqrt{(q_1 v_1)^2+(q_2 v_2)^2+(q_\Phi v_\Phi)^2}$ is the axion decay constant. Here, we note that the PQ charges are constrained to $pq_\Phi-q_1+q_2=0$ for the PQ invariance. We also can rewrite the QCD axion in terms of the orthogonal set of fields as
\bea
a =\frac{1}{v f_a}\bigg[(q_1 v^2_1+q_2 v^2_2) G+ \frac{p q_\Phi v_1 v_2  }{ v K}\,{\bar a} \bigg].
\eea
Therefore, in unitary gauge with $G=0$, the QCD axion is identical to the orthogonal scalar ${\bar a}$. In this case, for $v_\Phi\gg v_1, v_2$, we can approximate $K\simeq \frac{p v_1 v_2}{v^2 v_\Phi}$ and $f_a\simeq q_\Phi v_\Phi$, so $a\simeq {\bar a}$.
In the decoupling limit of the heavy pseudo-scalar with $A=0$, the QCD axion is dominated by the angular mode of the PQ field as $a\simeq a_\Phi$ from eq.~(\ref{QCDaxion}).

For $v_\Phi\gg v_1, v_2$, the PQ violating terms only with the PQ fields affect the axion quality problem most. So, we can take $n=0$ terms and replace $\rho$ by its VEV in the first line in the PQ violating potential in eq.~(\ref{PQV}). Then, for $a\simeq a_\Phi$ and $f_a\simeq q_\Phi v_\Phi$, the PQ violating terms at the order of $f^l_a$ are approximated to
\bea
V_{\rm PQV}\simeq \sum_{k=0}^{[l/2]} \frac{|c_{0,l,k}|  \, f^l_a }{2^{l/2} M_P^{l-4} q^l_\Phi}\cos\Big[(l-2k)\frac{q_\Phi a}{f_a}+A_{0,l,k}\Big]. \label{PQVpot}
\eea

In the presence of the PQ anomalies, we get the effective gluon couplings, as follows, 
\bea
{\cal L}_{\rm gluons}=\frac{g^2_s}{32\pi^2} \Big({\bar\theta}+\xi\frac{a}{f_a}\Big) G^a_{\mu\nu}{\tilde G}^{a\mu\nu}
\eea
where $\xi$ is the PQ anomaly coefficient, which is set to $\xi=1$ in KSVZ models and $\xi=k_G=6q_\Phi$ in DFSZ models.
Here, we can check the PQ anomaly coefficient $\xi$ explicitly for Type II and Y models in Table~2.
After PQ and electroweak symmetries are broken, the effective Yukawa interactions for the pseudo-scalars, $a_1\equiv \theta_1 v_1, a_2\equiv \theta_2 v_2$, appearing in the Higgs fields, are given by
\bea
{\cal L}_{{\rm eff},\theta_{1,2}}=-\frac{1}{\sqrt{2}}v_1 Y_d e^{i\theta_1} {\bar d}_L d_R -\frac{1}{\sqrt{2}} v_2 Y_u e^{-i\theta_2} {\bar u}_L u_R +{\rm h.c.}
\eea
Then, after making chiral rotations of quarks for three generations by $u_L\to e^{-i\theta_2/2} u_L, u_R\to e^{i\theta_2/2} u_R,  d_L\to e^{i\theta_1/2} d_L$ and  $d_R\to e^{-i\theta_1/2} d_R$ \cite{kim-review}, we obtain the anomalous shift of the Lagrangian as
\bea
\Delta {\cal L}=3\times \frac{\theta_1-\theta_2}{32\pi^2}\, G^a_{\mu\nu}{\tilde G}^{a\mu\nu}.
\eea
Here, from eq.~(\ref{pseudoscalar}), we can rewrite $\theta_1-\theta_2=\frac{a_1}{v_1}-\frac{a_2}{v_2}=\frac{pa_\Phi}{v_\Phi}-\frac{A}{N}$.
Thus, setting $A=0$ and using $a_\Phi\simeq a$ and $f_a\simeq q_\Phi v_\Phi$, we obtain $\theta_1-\theta_2\simeq \frac{p q_\Phi a}{f_a}$, so the anomaly coefficient $\xi$ becomes $\xi=3pq_\Phi$.

Then, after the QCD phase transition, there appears an extra contribution to the axion potential, in the following form,
\bea
\Delta V_E =-\Lambda^4_{\rm QCD} \cos\Big({\bar\theta}+\xi\frac{a}{f_a}\Big). \label{QCDpot}
\eea
 After the radial mode settles down to the minimum of the potential, i.e. $\langle\rho\rangle\simeq v_\Phi$, from eqs.~(\ref{PQVpot}) and (\ref{QCDpot}), the effective potential for the axion after the QCD phase transition is given by
\bea
V_{\rm eff}(a) =-\Lambda^4_{\rm QCD} \cos\Big({\bar\theta}+\xi\frac{a}{f_a}\Big)+M^4_P\bigg(\frac{f_a}{\sqrt{2}q_\Phi M_P}\bigg)^l \,\sum_{k=0}^{[l/2]} |c_{0,l,k}| \cos\Big((l-2k)\frac{q_\Phi a}{f_a}+A_{0,l,k}\Big). \label{axionpot}
\eea

In order to solve the strong CP problem by the axion, the axion potential needs to relax the effective $\theta$ term dynamically to satisfy the EDM bound,
\bea
|\theta_{\rm eff}|=\bigg|{\bar\theta}+\xi\frac{\langle a\rangle}{f_a}\bigg|<10^{-10}. \label{EDMbound}
\eea
Then, from the minimization of the effective potential for the axion in eq.~(\ref{axionpot}), namely, $\frac{dV_{\rm eff}}{da}=0$, we obtain
\bea
a_{\rm phys}\equiv a+\frac{f_a}{\xi} {\bar\theta} \simeq \frac{f^{l-1}_a}{2^{l/2}M^{l-4}_P q^{l-1}_\Phi m^2_a}\, \sum_{k=0}^{[l/2]} |c_{0,l,k}|  (l-2k)\sin\Big( A_{0,l,k}-\frac{l-2k}{\xi}\,q_\Phi{\bar\theta}\Big) \label{axionVEV}
\eea
where $m^2_a=\frac{\xi^2}{f^2_a}\,\Lambda^4_{\rm QCD}$ is the squared mass for the axion due to QCD only, and we assumed $(l-2k) q_\Phi \frac{a_{\rm phys}}{f_a}\ll 1$ and $(l-2k)^2|c_{0,l,k}|f^{l-2}_a/(2^{l/2}M^{l-4}_P q^{l-2}_\Phi)\cos\big( A_{0,l,k}-\frac{l-2k}{\xi}\,q_\Phi{\bar\theta}\big)\lesssim m^2_a$ for all $k$. 

As a result, from eq.~(\ref{axionVEV}) with eq.~(\ref{EDMbound}), we can solve the strong CP problem if
\bea
\frac{\xi f^{l-2}_a}{2^{l/2}M^{l-4}_P q^{l-1}_\Phi m^2_a}\, \sum_{k=0}^{[l/2]} |c_{0,l,k}| (l-2k)\sin\Big( A_{0,l,k}-\frac{l-2k}{\xi}\,q_\Phi{\bar\theta}\Big) < 10^{-10}.
\eea
Unless there is a cancellation between various contributions at the same order, each term in the PQ violating potential  at the order of $f^l_a$ is constrained by
\bea
\bigg(\frac{f_a}{M_P}\bigg)^{l}\lesssim \frac{2^{l/2}\xi \, q^{l-1}_\Phi}{(l-2k)|c_{0,l,k}|} \bigg(\frac{\Lambda_{\rm QCD}}{M_P}\bigg)^4\times 10^{-10}. \label{axionquality}
\eea 
As compared to the KSVZ models \cite{PQpole}, there is a mild dependence on the charge of the PQ field $q_\Phi$, but the order of magnitude estimation of the axion quality problem remains the same.
For a given axion decay constant, we can set a bound on the order of the PQ violating potential. For instance, choosing $f_a=10^{12}(10^8)\,{\rm GeV}$, $k=0$, $|c_{0,l,0}|={\cal O}(1)$ and $\xi=k_G=3pq_\Phi=6$ in Type II and Y models in Table~\ref{Anomalies},  we need $l\gtrsim 13(8)$ for the axion quality.
As will be discussed later, the bound from the CMB normalization leads to $3^{l/2}|c_{0,l,k}|\lesssim 10^{-10}$ for all $k$, so it is sufficient to take $l\gtrsim 11(7)$ for $f_a=10^{12}(10^8)\,{\rm GeV}$, $\xi=6$ and $|c_{0,l,k}|\lesssim 10^{-12}$.

\section{PQ inflation and predictions}

We derive the effective theory for a single-field slow-roll inflation. The first case is that the mixture of the radial components of PQ and neutral Higgs fields is responsible for the inflaton. The second case is that the pure radial component of the PQ field is the inflaton. We check the stability of the non-inflaton directions during inflation. Then, focusing on the pure PQ inflation, we present the inflatonary predictions and the initial condition for the axion velocity at the end of inflation due to the PQ violating potential.

\subsection{Effective theory for PQ-Higgs mixed inflation}

For general inflation dynamics in DFSZ models, the Higgs fields can participate to the inflationary dynamics, so their background field values are nonzero, i.e. $h_1\neq 0, h_2\neq 0$. 
In this case, slow-roll inflaton takes place near the pole of the kinetic term, that is, $\chi^2\equiv \rho^2+h^2_1+h^2_2\to 6$. On the other hand, the orthogonal directions to the inflaton, denoted as $\tau_1=h_1/\rho$ and $\tau_2=h_2/\rho$, must be stabilized \cite{higgsportal,Choi:2019osi}. 
Then, noting $\rho^2=\chi^2/(1+\tau^2_1+\tau^2_2)$,  we can rewrite the kinetic terms in eq.~(\ref{Ein2}) for $\chi, \tau_1,\tau_2$ and the angular modes, as follows,
 \bea
 \frac{\mathcal{L}_{\mathrm{kin},E}}{\sqrt{-g_E}} &=& -\frac{1}{2}R  + \frac{ \left(\partial_\mu( \tau_1\rho)\right)^2 + \left(\partial_\mu (\tau_2\rho)\right)^2 + (\tau_1\rho)^2\left(\partial_\mu \eta_1\right)^2 + (\tau_2\rho)^2\left(\partial_\mu \eta_2\right)^2 + \left(\partial_\mu \rho\right)^2 + \rho^2\left(\partial_\mu \theta\right)^2}{2(1 - \frac{1}{6}\chi^2)} \nonumber\\
&&\quad+ \frac{3}{4} \frac{\Big(\partial_\mu\left(\frac{1}{6}\chi^2\right)\Big)^2}{( 1- \frac{1}{6}\chi^2)^2} \nonumber \\
&=& -\frac{1}{2}R  +\frac{1}{2} \frac{(\partial_\mu\chi)^2}{(1- \frac{1}{6}\chi^2)^2} +\frac{\chi^2}{2( 1- \frac{1}{6}\chi^2)}\frac{1}{(1+\tau^2_1+\tau^2_2)}\bigg[(\partial_\mu\theta)^2+\tau^2_1 (\partial_\mu\eta_1)^2+\tau^2_2 (\partial_\mu\eta_2)^2 \bigg]  \nonumber \\
&&+\frac{\chi^2}{2( 1- \frac{1}{6}\chi^2)} \frac{1}{(1+\tau^2_1+\tau^2_2)^2} \bigg[(\partial_\mu\tau_1)^2+(\partial_\mu\tau_2)^2 +(\tau_2\partial_\mu\tau_1-\tau_1\partial_\mu \tau_2)^2\bigg].
 \eea
Introducing the canonical inflaton by
\bea
\chi=\sqrt{6} \tanh\bigg(\frac{\phi}{\sqrt{6}}\bigg),
\eea
we can simplify the above kinetic terms as
\bea
 \frac{\mathcal{L}_{\mathrm{kin},E}}{\sqrt{-g_E}} &=& -\frac{1}{2}R  +\frac{1}{2} (\partial_\mu\phi)^2 +\frac{3\sinh^2\bigg(\frac{\phi}{\sqrt{6}}\bigg) }{(1+\tau^2_1+\tau^2_2)} \bigg[(\partial_\mu\theta)^2+\tau^2_1 (\partial_\mu\eta_1)^2+\tau^2_2 (\partial_\mu\eta_2)^2 \bigg]  \nonumber \\
 &&+\frac{3\sinh^2\bigg(\frac{\phi}{\sqrt{6}}\bigg) }{(1+\tau^2_1+\tau^2_2)^2}  \bigg[(\partial_\mu\tau_1)^2+(\partial_\mu\tau_2)^2 +(\tau_2\partial_\mu\tau_1-\tau_1\partial_\mu \tau_2)^2\bigg].
\eea

For $\langle\tau_1\rangle\neq 0$ and $\langle\tau_2\rangle\neq 0$, there is a nonzero PQ invariant potential for one of the angular modes, ${\tilde A}\equiv \eta_2-\eta_1+p\theta$, which is stabilized at either $0$ or $\pi$.  Moreover, $\langle \tau^2_1\rangle\eta_1+\langle \tau^2_2\rangle\eta_2$ is eaten by the $Z$ boson as in the vacuum in Section 3.3. In unitary gauge, we can choose $\langle \tau^2_2\rangle\eta_2=-\langle \tau^2_1\rangle\eta_1$, so the heavy angular direction becomes ${\tilde A}\equiv -(1+\langle \tau^2_1\rangle/\langle \tau^2_2\rangle)\eta_1+p\theta$.
Thus, the Lagrangian for the angular modes become
\bea
 \frac{\mathcal{L}_{\mathrm{angles},E}}{\sqrt{-g_E}} =\frac{3\sinh^2\bigg(\frac{\phi}{\sqrt{6}}\bigg) }{(1+\tau^2_1+\tau^2_2)} \bigg[ (\partial_\mu\theta)^2+\frac{\tau^2_1\langle \tau^4_2\rangle+\tau^2_2\langle \tau^4_1\rangle}{(\langle \tau^2_1\rangle+\langle \tau^2_2\rangle)^2} (\partial_\mu {\tilde A}- p\partial_\mu\theta)^2 \bigg]- V_{\rm angles}({\tilde A}),
\eea
with
\bea
V_{\rm angles}({\tilde A})=\frac{\kappa_p\tau_1\tau_2\chi^{p+2}}{2 (1+\tau^2_1+\tau_2^2)^{(p+2)/2}}\,\cos {\tilde A}.
\eea
Then, we can stabilize ${\tilde A}=0$ or $\pi$, depending on $\lambda_{12\Phi}<0$ or $\lambda_{12\Phi}>0$.
For instance, for $\langle\tau_1\rangle=\langle\tau_2\rangle$ and $p=2$, the effective mass for ${\tilde A}$ becomes
\bea
m^2_{{\tilde A},{\rm eff}} =\frac{6|\kappa_2| \sinh^2\bigg(\frac{\phi}{\sqrt{6}}\bigg) }{(1+2\langle\tau^2_1\rangle)\cosh^4\bigg(\frac{\phi}{\sqrt{6}}\bigg)},
\label{Aeffmass}
\eea
which is approximated to $m^2_{{\tilde A},{\rm eff}}\simeq \frac{6|\kappa_2|}{1+2\langle\tau^2_1\rangle}\,\sqrt{\frac{3}{4}\epsilon}$ for $\phi\gg \sqrt{6}$ during inflation.
So, the angular mode ${\tilde A}$ is decoupled during inflation, if $m^2_{{\tilde A},{\rm eff}} \gg H^2_I\simeq 3\lambda_\Phi$.

Furthermore, after stabilizing ${\tilde A}$, we can recast the scalar potential in terms of $\chi, \tau_1,\tau_2$ into
{\small
\bea
V_{\rm PQ}&=& \frac{\chi^4}{4(1+\tau^2_1+\tau^2_2)^2}\bigg(\lambda_1\tau^4_1+\lambda_2\tau^4_2+(\lambda_3+\lambda_4)\tau^2_1\tau^2_2 +\lambda_\Phi+\lambda_{1\Phi} \tau^2_1+\lambda_{2\Phi} \tau^2_2 \nonumber \\
&&-\frac{2|\kappa_p|\chi^{p-2} \tau_1\tau_2}{(1+\tau^2_1+\tau^2_2)^{(p-2)/2}} \bigg), \\
V_{\rm PQV}&=&\sum_{n,l}\sum_{k=0}^{[l/2]} \frac{|c_{n,l,k}|(\tau_1\tau_2)^n \chi^{2n+l}}{2^{n+l/2} M_P^{2n+l-4}(1+\tau^2_1+\tau^2_2)^{(2n+l)/2}}\,\cos\Big[(l-2n-2k)\theta+A_{n,l,k}+n \alpha\Big],
\eea
}
with $\alpha=0$ or $\pi$, depending on whether ${\tilde A}$ is stabilized at $0$ or $\pi$.
 
 As a result, after stabilizing ${\tilde A}$, the effective Lagrangian for the general inflation is given by
{\small
\bea
 \frac{\mathcal{L}_{\mathrm{eff},E}}{\sqrt{-g_E}} &=& -\frac{1}{2}R  +\frac{1}{2} (\partial_\mu\phi)^2+\frac{3\sinh^2\Big(\frac{\phi}{\sqrt{6}}\Big) }{(1+\tau^2_1+\tau^2_2)} \bigg(1+\frac{p^2\tau^2_1\langle \tau^4_2\rangle+p^2\tau^2_2\langle \tau^4_1\rangle}{(\langle \tau^2_1\rangle+\langle \tau^2_2\rangle)^2}\bigg)(\partial_\mu\theta)^2 \nonumber \\
 &&+\frac{3\sinh^2\Big(\frac{\phi}{\sqrt{6}}\Big) }{(1+\tau^2_1+\tau^2_2)^2}  \bigg[(\partial_\mu\tau_1)^2+(\partial_\mu\tau_2)^2 +(\tau_2\partial_\mu\tau_1-\tau_1\partial_\mu \tau_2)^2\bigg]- V_E(\phi,\theta,\tau_1,\tau_2). \label{Leffgen}
\eea
}
Here, $V_E=V_{\rm PQ}+V_{\rm PQV}$, and  the effective scalar potentials are written as factorizable forms for $p=2$,
\bea
V_{\rm PQ}=U(\phi) \cdot W(\tau_1,\tau_2), \label{VPQgen}
\eea
with 
\bea
U(\phi)=9 \lambda_\Phi \tanh^4\Big(\frac{\phi}{\sqrt{6}}\Big)
\eea 
and
\bea
W= \frac{1}{\lambda_\Phi(1+\tau^2_1+\tau^2_2)^2}\Big(\lambda_1\tau^4_1+\lambda_2\tau^4_2+(\lambda_3+\lambda_4)\tau^2_1\tau^2_2+\lambda_\Phi+\lambda_{1\Phi} \tau^2_1+\lambda_{2\Phi} \tau^2_2-2 |\kappa_2| \tau_1\tau_2\Big), \label{Wfunc}
\eea
and
\bea
V_{\rm PQV}= \sum_{n,l}\sum_{k=0}^{[l/2]} \frac{3^{n+l/2}|c_{n,l,k}|}{ M_P^{2n+l-4}}\, \frac{(\tau_1\tau_2)^n  \tanh^{2n+l}\Big(\frac{\phi}{\sqrt{6}}\Big)}{(1+\tau^2_1+\tau^2_2)^{(2n+l)/2}}\,\cos\Big[(l-2n-2k)\theta+A_{n,l,k}+n \alpha\Big]. \label{VPQVgen}
\eea
However, for $p=1$, the PQ-invariant effective scalar potential is not of factorizable form, but instead it has a correction to the potential for $\phi$, proportional to the Higgs VEVs, as follows,
\bea
V_{\rm PQ}= U(\phi) \Big(A(\tau_1,\tau_2)+P(\phi)B(\tau_1,\tau_2) \Big),
\eea
with
\bea
A(\tau_1,\tau_2) &=&\frac{1}{\lambda_\Phi(1+\tau^2_1+\tau^2_2)^2}\Big(\lambda_1\tau^4_1+\lambda_2\tau^4_2+(\lambda_3+\lambda_4)\tau^2_1\tau^2_2+\lambda_\Phi+\lambda_{1\Phi} \tau^2_1+\lambda_{2\Phi} \tau^2_2\Big), \\
B(\tau_1,\tau_2) &=&\frac{\tau_1\tau_2}{ (1+\tau^2_1+\tau^2_2)^{(p+2)/2} },  \label{Afunc} \\
P(\phi)&=& -2 \cdot 6^{(p-2)/2}\frac{|\kappa_p|}{\lambda_\Phi} \tanh^{p-2}\Big(\frac{\phi}{\sqrt{6}}\Big).
\eea

Now we check the stability for $\tau_1$ and $\tau_2$. For $\lambda_1=\lambda_2$ and $\lambda_{1\Phi}=\lambda_{2\Phi}$, the Lagrangian is symmetric under the exchange of $\tau_1$ and $\tau_2$, so are the VEVs, that is, $\langle\tau_1\rangle=\langle\tau_2\rangle$. Otherwise, $\langle\tau_1\rangle\neq \langle\tau_2\rangle$.
The minimization conditions, $\frac{\partial V_{\rm PQ}}{\partial \tau_1}=0$ and $\frac{\partial V_{\rm PQ}}{\partial \tau_2}=0$, are satisfied for $\langle\tau_1\rangle=\langle\tau_2\rangle=0$, which is the case for pure PQ inflation in the next subsection. 
But, $\langle\tau_1\rangle\neq 0, \langle\tau_2\rangle=0$ or $\langle\tau_1\rangle=0,\langle\tau_2\rangle\neq 0$ are not the minimum of the potential for $\kappa_p\neq 0$, due to the effective tadpole term for either directions \cite{multifield}. Thus, if $\langle\tau_1\rangle\neq 0$, we need $\langle\tau_2\rangle\neq 0$, and vice versa. The values of $\langle\tau_1\rangle$ and $\langle\tau_2\rangle$ are bounded because 
\bea
\chi=\rho\,(1+\tau^2_1+\tau^2_2)^{1/2}=\sqrt{6}  \tanh\bigg(\frac{\phi}{\sqrt{6}}\bigg)\leq \sqrt{6}.
\eea
The general solutions to  $\frac{\partial V_{\rm PQ}}{\partial \tau_1}=0$ and $\frac{\partial V_{\rm PQ}}{\partial \tau_2}=0$ are involved, but the stable solutions for inflation exist, as far as the Hessian of the squared mass matrix for $\tau_1, \tau_2$ in the vacuum is positive and the inflation vacuum energy is positive for $W(\langle \tau_1\rangle, \langle\tau_2\rangle)>0$ for $p=2$ and $A(\langle\tau_1\rangle,\langle\tau_2\rangle) +P(\phi)B(\langle\tau_1\rangle,\langle\tau_2\rangle)>0$ for $p=1$ \cite{higgsportal}.

After $\tau_1$ and $\tau_2$ are stabilized, the effective kinetic term for $\theta$ in eq.~(\ref{Leffgen}) becomes
\bea
\frac{3\sinh^2\Big(\frac{\phi}{\sqrt{6}}\Big) }{(1+\langle\tau^2_1\rangle+\langle\tau^2_2\rangle)} \bigg(1+\frac{p^2\langle\tau^2_1\rangle\langle\tau^2_2\rangle}{\langle\tau^2_1\rangle+\langle\tau^2_2\rangle}\bigg)(\partial_\mu\theta)^2.
\eea
In particular, for $\langle\tau_1\rangle=\langle\tau_2\rangle$, the above kinetic term for $\theta$ becomes simplified to
\bea
\bigg(\frac{2+p^2\langle\tau^2_1\rangle}{2(1+2\langle\tau^2_1\rangle)}\bigg)3\sinh^2\bigg(\frac{\phi}{\sqrt{6}}\bigg)(\partial_\mu\theta)^2,
\eea
which is of the same form as in KSVZ models for $p=2$ \cite{PQpole}. 
Similarly, the PQ invariant potential eq.~(\ref{VPQgen}) become functions of $\phi$ only and the PQ violating potential in eq.~(\ref{VPQVgen}) are functions of $\phi$ and $\theta$ only. 

We remark that the effective vacuum energy during inflation is controlled by $W(\tau_1,\tau_2)$ in eq.~(\ref{Wfunc}) or $A(\tau_1,\tau_2)$ in eq.~(\ref{Afunc}) with the VEVs of $\tau_1$ and $\tau_2$. Then, for  $\langle\tau_1\rangle\neq 0$ or $\langle\tau_2\rangle\neq 0$,  the quartic couplings for two Higgs doublets contribute to the effective vacuum energy. So, the CMB normalization requires a suppressed vacuum energy during inflation as will be discussed later, we need very small $\langle\tau_1\rangle$ and $\langle\tau_2\rangle$ or the quartic couplings for two Higgs doublets must run to very small values during inflation \cite{Higgspole}. In both cases, we would need nontrivial relations between the running couplings for small Higgs VEVs or vanishing beta functions during inflation. Thus, as will be discussed in the next subsection, we focus on the case for pure PQ inflation where the Higgs VEVs are stabilized at the origin during inflation.

\subsection{Effective theory for pure PQ inflation}

We take the inflation direction along the radial mode of the PQ field, so the background field values for two Higgs doublets are set to $\langle h_1\rangle=\langle h_2\rangle=0$ or $\langle\tau_1\rangle=\langle\tau_2\rangle=0$. In this case, the angular modes also vanish in the polar representations in eq.~(\ref{polarrep}). So, instead we need to include the imaginary partners of the neutral Higgs scalars in the Cartesian representations, instead of $\eta_1$ and $\eta_2$ in eq.~(\ref{polarrep}), that is, $\frac{1}{\sqrt{2}} h_i \,e^{i\theta_1}\to \frac{1}{\sqrt{2}} (h_i +i{\tilde\eta}_i)$ with $i=1,2$. Then, it is obvious that the imaginary parts of the neutral Higgs scalars receive the same masses as for  ${\bar \tau}_1, {\bar \tau_2}$, due to the symmetry of the potential respecting the SM gauge symmetry during inflation. We can still use the polar representation for the PQ field as in eq.~(\ref{polarrep}), because the radial mode of the PQ field is nonzero during inflation so the angular mode of the PQ field is kept. 

For the pure PQ inflation, we get the effective Lagrangian for describing the evolution of the background, that is, the inflaton and the angular mode associated with the PQ field, as 
\bea
\frac{\mathcal{L}_{\mathrm{bkg},E}}{\sqrt{-g_E}}=- \frac{1}{2}R+\frac{1}{2}(\partial_\mu\phi)^2+3\sinh^2\bigg(\frac{\phi}{\sqrt{6}}\bigg) \left(\partial_\mu\theta\right)^2-V_E(\phi,\theta) \label{Lbkg}
\eea
where the inflaton potential in Einstein frame is given by $V_E=V_{\rm PQ}(\phi)+V_{\rm PQV}(\phi,\theta) $, with
\bea
V_{\rm PQ}(\phi)&=&9 \lambda_\Phi  \tanh^4\bigg(\frac{\phi}{\sqrt{6}}\bigg)=U(\phi),  \\
V_{\rm PQV}(\phi,\theta) &=&\sum_{k=0}^{[l/2]}\frac{3^{l/2}|c_{0,l,k}|}{M_P^{l-4}}\,  \tanh^l\bigg(\frac{\phi}{\sqrt{6}}\bigg)  \cos((l-2k)\theta+A_{0,l,k}). 
\label{PQpure-PQV}
\eea
Here, the background field values for the Higgs-dependent terms in the Lagrangian were set to zero and we kept all the terms contributing to $\rho^l$ in the PQ violating potential.
The inflaton dynamics is governed dominantly by the quartic interaction in $V_{\rm PQ}$
whereas the angular mode $\theta$ associated with the PQ field receives a nonzero velocity due to $V_{\rm PQV}$.  The resulting effective Lagrangian for the PQ inflation in DFSZ models takes the same form as in KSVZ models \cite{PQpole}. 

The effective Lagrangian for the real parts of the Higgs perturbations, ${\bar h}_1, {\bar h_2}$, or ${\bar\tau}_1,{\bar\tau}_2$, up to quadratic terms, are given by
\bea
\frac{\mathcal{L}_{\mathrm{pert},E}}{\sqrt{-g_E}}= 3\sinh^2\bigg(\frac{\phi}{\sqrt{6}}\bigg) \Big((\partial_\mu {\bar \tau}_1)^2+(\partial_\mu {\bar \tau}_2)^2 \Big)- \frac{1}{2} m^2_{11} {\bar \tau}^2_1- \frac{1}{2} m^2_{22} {\bar \tau}^2_2 -m^2_{12} {\bar\tau}_1 {\bar \tau}_2
\eea
with
\bea
m^2_{ii} &=& \bigg(\frac{1}{2} \lambda_{i\Phi}-\lambda_\Phi\bigg) \rho^4 =18( \lambda_{i\Phi}-2\lambda_\Phi )\tanh^4 \bigg(\frac{\phi}{\sqrt{6}}\bigg), \quad i=1,2, \\
m^2_{12}&=&-\frac{1}{2} \kappa_p  \rho^{p+2} = -\frac{1}{2}\, 6^{p/2+1}\kappa_p  \tanh^{p+2} \bigg(\frac{\phi}{\sqrt{6}}\bigg).
\eea
Then, after diagonalizing the mass matrix for  ${\bar\tau}_1,{\bar\tau}_2$, we obtain the mass eigenvalues as
\bea
m^2_\pm =\frac{1}{4}\bigg[  \lambda_{1\Phi}+ \lambda_{2\Phi}-4\lambda_\Phi \pm \sqrt{( \lambda_{1\Phi}- \lambda_{2\Phi})^2+4\rho^{2p-4}\kappa^2_p}\bigg]\rho^4,
\eea 
which are positive definite as far as
\bea
( \lambda_{1\Phi}-2\lambda_\Phi)( \lambda_{2\Phi}-2\lambda_\Phi)> \rho^{2p-4}\kappa^2_p\simeq 6^{p-2}\kappa^2_p. \label{stability}
\eea
For instance, for $p=2$,  the effective masses for the canonical Higgs perturbations are
\bea
m^2_{\pm,{\rm eff}}&=&\frac{m^2_\pm }{ 6\sinh^2\bigg(\frac{\phi}{\sqrt{6}}\bigg) } \nonumber \\
&\simeq&\frac{3}{2} \bigg[  \lambda_{1\Phi}+ \lambda_{2\Phi}-4\lambda_\Phi \pm \sqrt{( \lambda_{1\Phi}- \lambda_{2\Phi})^2+ 4\kappa^2_2}\bigg] \cdot \frac{\sinh^2 \bigg(\frac{\phi}{\sqrt{6}}\bigg)}{\cosh^4 \bigg(\frac{\phi}{\sqrt{6}}\bigg)}, \label{effmasses}
\eea
where $\sinh^2 \big(\frac{\phi}{\sqrt{6}}\big)/\cosh^4 \big(\frac{\phi}{\sqrt{6}}\big)\simeq 4 e^{-2\phi/\sqrt{6}}\simeq \sqrt{\frac{3}{4}\epsilon}$ during inflation. 
So, the Higgs directions are decoupled during inflation, as far as $m^2_{\pm,{\rm eff}} \gg H^2_I\simeq 3\lambda_\Phi$.

\subsection{Slow-roll inflation and axion rotation}

After stabilizing the field directions other than $\phi$ and $\theta$, we consider the slow-roll inflation and the initial condition for the axion velocity at the end of inflation. 
As discussed in the previous subsection, we can stabilize the Higgs fields at the origin due to their large positive effective masses for which  the pure PQ inflation is realized by $\phi$. In this case, the axion field $\theta$ receives a nonzero initial velocity due to the PQ violating potential.

From the effective Lagrangian in eq.~(\ref{Lbkg}), we obtain the equation of motion for the radial and angular modes of the PQ field in the following,
\bea
&&{\ddot\phi}+3H {\dot \phi} -\sqrt{6}\sinh\Big(\frac{\phi}{\sqrt{6}}\Big) \cosh\Big(\frac{\phi}{\sqrt{6}}\Big) \, {\dot\theta}^2=-\frac{\partial V_E}{\partial \phi}, \label{eom-phi} \\
&&6\sinh^2\Big(\frac{\phi}{\sqrt{6}}\Big)\bigg[{\ddot\theta}+ 3H {\dot \theta} + \frac{2}{\sqrt{6} } \coth\Big(\frac{\phi}{\sqrt{6}}\Big)\, {\dot\phi}\,{\dot\theta}\bigg] = -\frac{\partial V_E}{\partial\theta}.  \label{eom-theta}
\eea
Here, the Hubble parameter is determined by
\bea
H^2=\frac{1}{3}\, \bigg(\frac{1}{2}(\partial_\mu \phi)^2+ 3\sinh^2\Big(\frac{\phi}{\sqrt{6}}\Big)\,(\partial_\mu\theta)^2+ V_E\bigg).
\eea
which is approximate to $H^2\simeq \frac{V_E}{3}\simeq 3 \lambda_\Phi$ during inflation. 

For a slow-roll inflation with ${\ddot\phi}\ll H {\dot\phi}$ and ${\dot\phi}\ll H$ as well as ${\ddot\theta}\ll H {\dot\theta}$ and ${\dot\theta}\ll H$, 
we simply approximate eqs.~(\ref{eom-phi}) and (\ref{eom-theta}) as
\bea
{\dot\phi}&\simeq& -\frac{1}{3H} \frac{\partial V_E}{\partial \phi}=-\sqrt{2\epsilon_\phi}\, M_P H, \label{slow-phi} \\
{\dot\theta}&\simeq&  -\frac{1}{3H} \, \frac{\frac{\partial V_E}{\partial\theta}}{6M^2_P\sinh^2\big(\frac{\phi}{\sqrt{6}M_P}\big)}=-\frac{\sqrt{2\epsilon_\theta} \,H}{6\sinh^2\big(\frac{\phi}{\sqrt{6}M_P}\big)} \label{slow-theta}
\eea
where $\epsilon_\phi, \epsilon_\theta$ are the slow-roll parameters for the radial and angular modes, respectively.
Then, the radial mode derives a slow-roll inflation while the dynamics of the angular mode is subdominant during inflation. 
As a result, we can set the initial kinetic misalignment of the axion by using eq.~(\ref{slow-theta}) at the end of inflation \cite{PQpole}, giving rise to the PQ Noether charge at the end of inflation as
\bea
n_{\theta,{\rm end}}=6 \sinh^2\Big(\frac{\phi_{\rm end}}{\sqrt{6}}\Big) |{\dot\theta}_{\rm end}|\simeq  \sqrt{2\epsilon_{\theta,{\rm end}}} \, H_{\rm end}.
\label{PQcharge}
\eea
Here, $\epsilon_{\theta,{\rm end}}, H_{\rm end}, \phi_{\rm end}$ are the quantities evaluated at the end of inflation.

We assume the PQ invariant potential to be dominant for the pure PQ inflation, so the inflaton potential is approximately given by
\bea
V_E(\phi)\simeq V_I \bigg[  \tanh^4\Big(\frac{\phi}{\sqrt{6}}\Big)\bigg], \label{inflation1}
\eea
with $V_I\equiv 9\lambda_\Phi$.
Then, the slow-roll parameters during inflation are given by
\bea
\epsilon 
&=& \frac{16}{3}  \bigg[  \sinh\Big(\frac{2\phi}{\sqrt{6}}\Big)\bigg]^{-2} ,  \label{ep} \\
\eta 
&=&-\frac{8}{3} \bigg[  \cosh\Big(\frac{2\phi}{\sqrt{6}}\Big)-4 \bigg]  \bigg[\sinh\Big(\frac{2\phi}{\sqrt{6}}\Big)\bigg]^{-2}. \label{eta}
\eea
The number of efoldings is 
\bea
N&=&\frac{1}{M_P} \int^{\phi_*}_{\phi_e} \frac{ {\rm sgn} (V'_E)d\phi}{\sqrt{2\epsilon}} \nonumber \\
&=&\frac{3}{8}\, \bigg[ \cosh\Big(\frac{2\phi_*}{\sqrt{6}}\Big)- \cosh\Big(\frac{2\phi_e}{\sqrt{6}}\Big)  \bigg] \label{efold}
\eea
where $\phi_*, \phi_e$ are the values of the radial mode at horizon exit and the end of inflation, respectively. 
As a result, using  eqs.~(\ref{ep}), (\ref{eta}) and (\ref{efold}) and $N\simeq \frac{3}{8}\,  \cosh\Big(\frac{2\phi_*}{\sqrt{6}M_P}\Big)$ for $\phi_*\gg \sqrt{6}$ during inflation, we obtain the slow-roll parameters at horizon exit in terms of the number of efoldings as
\bea
\epsilon_* \simeq  \frac{3}{4\big(N^2-\frac{9}{64}\big)},  \quad
\eta_* \simeq \frac{3-2N}{2\big(N^2-\frac{9}{64}\big)}. \label{slowroll}
\eea
Thus, we get the spectral index in terms of the number of efoldings and the tensor-to-scalar ratio, as follows, 
\bea
n_s&=& 1-\frac{4N+3}{2\big(N^2-\frac{9}{64}\big)}, \label{sindex} \\
r&=&16\epsilon_* =  \frac{12}{N^2-\frac{9}{64}}. \label{ratio}
\eea
We note that if a higher order PQ invariant potential such as $\beta_m |\Phi|^{2m}$ with $m>2$ is introduced instead of the quartic potential $\lambda_\Phi |\Phi|^4$, the inflaton potential is changed to the form, $V_E(\phi)\propto \tanh^{2m}(\phi/\sqrt{6})$, but the inflationary predictions are insensitive to the value of $m$ \cite{PQpole}.

The inflationary predictions in the PQ pole inflation are  $n_s=0.966$ and $r=0.0033$ for $N=60$, which are consistent with the observations, namely, $n_s=0.967\pm 0.0037 $ \cite{planck} and  $r<0.036$ \cite{keck}.
On the other hand, from the CMB normalization, $A_s=\frac{1}{24\pi^2} \frac{V_I}{\epsilon_*}=2.1\times 10^{-9}$ \cite{planck},  we need to set the quartic coupling for the PQ field  to $\lambda_\Phi=1.1\times 10^{-11}$ during inflation. 
Moreover, the PQ invariant mass term is bounded by
$|\mu_\Phi|<1.4\times 10^{13}\,{\rm GeV}$,
and there are similar bounds on the PQ violating terms as
\bea
V_{\rm PQV}= 3^{l/2}\sum_{k=0}^{[l/2]} |c_{0,l,k}| \cos\Big((l-2k)\theta_i+A_{0,lk}\Big) <1.0\times 10^{-10}. \label{PQVbound}
\eea
We note that the slow-roll parameter $\epsilon_\theta$ appearing in the PQ Noether charge in eq.~(\ref{PQcharge}) depends on the PQ violating potential in eq.~(\ref{PQpure-PQV}), so it is smaller than unity as far as the coefficients of the PQ violating potential at the order $l$ of the PQ field satisfy $3^{l/2}|c_{0,l,k}| <1.0\times 10^{-10}$ for each $k$ from eq.~(\ref{PQVbound}).
In this case, as discussed in the end of Section 3.3, there is no axion quality problem as far as  the order of the PQ violating potential is given by $l\gtrsim 11(7)$ for $f_a=10^{12}(10^8)\,{\rm GeV}$, $\xi=6$ and $|c_{0,l,k}|\lesssim 10^{-12}$.

We remark on the multiple dynamics other than the radial component of the PQ field. In both the PQ-Higgs mixed inflation and the pure PQ inflation, it was shown that the field directions orthogonal to the inflaton field can receive masses much greater than the Hubble scale during inflation. But, if the masses of the non-inflaton fields are comparable to or smaller than the Hubble scale in a certain parameter space, we need to consider the dynamics of multiple fields during inflation, for instance, for the non-Gaussianity of CMB \cite{non-G}.
However, the detailed discussion on the dynamics of multiple fields and the non-Gaussianity is beyond the scope of our work, so we postpone it in a future work.

\section{Reheating}

In this section, we discuss the basics of the inflaton condensate after inflation and provide the results for the rates for inflaton decays and scatterings. Using the results, we determine the reheating temperature and axion dark radiation and comment on the condition for the PQ symmetry restoration after reheating.

\subsection{Inflaton condensate}

When inflation is driven by the radial mode of the PQ field, the inflaton field value at the end of inflation is given by $\phi_{\rm end}=\sqrt{\frac{3}{8}}\ln\big(\frac{1}{6}(35+8\sqrt{19})\big)\simeq 1.50$ from ${\ddot a}=0$, for which the inflaton energy density at the end of inflation is given by $\rho_{\rm end}=\frac{3}{2} V_E(\phi_{\rm end})$.  The post-inflationary potential of the inflaton for $|\Phi|\ll \sqrt{3}$ takes $V_E(\phi)\simeq \frac{1}{4}\lambda_\Phi \phi^4$, so the equation of state for the inflaton during reheating is radiative-like, i.e. $w_\phi=p_\phi/\rho_\rho=\frac{1}{3}$.
 
The inflaton condensate during reheating takes $\phi=\phi_0(t){\cal P}(t)$ where $\phi_0(t)$ is the amplitude of the inflaton, which is constant over one oscillation \footnote{The amplitude of the inflaton undergoes damping during the Hubble expansion as $\phi_0(t)=\phi_{\rm end}\sqrt{t_{\rm end}/t}$  in the case of $V_E(\phi)=\frac{1}{4}\lambda_\Phi \phi^4$ during reheating.}. Here, ${\cal P}(t) $ is the periodic function satisfying ${\dot{\cal P}}^2=\frac{2\rho_\phi}{\phi^2_0}(1-{\cal P}^4)$, which is given \cite{Greene:1997fu,Choi:2019osi} by
\bea
{\cal P}(t) ={\rm cn} \bigg({\bar\omega} t, \frac{1}{2}\bigg)
\eea
where ${\bar\omega}=\sqrt{\lambda_\Phi}\phi_0$ and ${\rm cn}(u,m)=\cos\varphi$ is the Jacobi cosine for $u=\int^\varphi_0 d\theta/\sqrt{1-m^2\sin^2\theta}$. We note that ${\bar\omega}$ is different from the angular frequency of oscillation $\omega$, which is given \cite{Higgspole,PQpole} by
\bea
\omega=m_\phi\sqrt{\frac{2\pi}{3}} \frac{\Gamma[\frac{3}{4}]}{\Gamma[\frac{1}{4}]}= \sqrt{2\pi \lambda_\Phi}  \, \frac{\Gamma[\frac{3}{4}]}{\Gamma[\frac{1}{4}]}\, \phi_0
\eea
where we used $m^2_\phi=V^{\prime\prime}_E(\phi_0)=3\lambda_\Phi \phi^2_0$, that is, $\omega=0.847\,{\bar\omega}$. 
As a result, we can make a Fourier expansion of the periodic function ${\cal P}$ by ${\cal P}(t)=\sum_{n=-\infty}^\infty {\cal P}_n \,e^{-in\omega t}$. Then, the first few nonzero coefficients are given by $2{\cal P}_1=0.9550, 2{\cal P}_3=0.04305, 2{\cal P}_5=0.001859$, etc.
Similarly, for the Fourier expansion of ${\cal P}^2$ as ${\cal P}^2(t)=\sum_{n=-\infty}^\infty ({\cal P}^2)_n e^{-in\omega t}$, the first few nonzero coefficients are given by $2({\cal P}^2)_2=0.4972, 2({\cal P}^2)_4=0.04289, 2({\cal P}^2)_6=0.002778$, etc.

\subsection{Reheating}

For the pure PQ inflation, reheating can take place due to the PQ invariant interactions to the Higgs fields and the Yukawa couplings to the RH neutrinos, 
\bea
{\cal L}_{\rm int}\supset-\sum_{i=1,2}\lambda_{i\Phi}|H_i|^2|\Phi|^2-2^{p/2-1}\kappa_p H^\dagger_1 H_2 \Phi^p-\frac{1}{2} y_N \overline{N^c_R} \Phi N_R+{\rm h.c.},
\eea
or Planck-scale suppressed derivative interactions to the Higgs fields of type $\phi^2 (\partial_\mu h_i)^2$, $\phi h_i (\partial_\mu\phi)(\partial^\mu h_i)$, as seen from the non-canonical kinetic terms of eq.~(\ref{Ein2}). In our work, we remind that it is necessary to introduce $\kappa_p H^\dagger_1 H_2 \Phi^p$ for realizing the consistent electroweak symmetry breaking and the QCD axion.

The PQ invariant interactions are responsible for reheating by the inflaton scattering with the Higgs-portal couplings including the $p=2$ term, namely, $\phi\phi\to H^\dagger_1 H_1, H^\dagger_2 H_2, H^\dagger_1 H_2, H^\dagger_2 H_1$ or the inflaton decay with the Higgs-portal coupling with $p=1$, such as $\phi\to H^\dagger_1 H_2, H^\dagger_2 H_1$, and the inflaton decay/scattering with the Yukawa couplings to the RH neutrinos, such as $\phi\to N_R  \overline{N^c_R}$, $\phi\phi\to N_R N_R$, and their hermitian conjugates.
Moreover, due to the self-interactions of the PQ field, we also need to consider the inflaton scattering into a pair of axions, i.e. $\phi\phi\to aa$. 

During reheating, the Boltzmann equations for the averaged  energy density for the inflaton and the radiation energy density $\rho_R$ are given by
\bea
{\dot\rho}_\phi +3(1+w_\phi) H\rho_\phi &=&-\Gamma_\phi(1+w_\phi) \rho_\phi, \\
{\dot\rho}_R + 4H \rho_R &=& \Gamma_\phi(1+w_\phi) \rho_\phi
\eea
where $\Gamma_\phi$ is the sum of inflaton decay or scattering rates, given by
\bea
\Gamma_\phi&=&\Gamma_{\phi\to N_RN_R, \overline {N_R}\, \overline {N_R}}+ \Gamma_{\phi\phi\to N_R N_R, \overline {N_R}\, \overline {N_R}} \nonumber \\
&&+\Gamma_{\phi\phi\to H^\dagger_i H_i}+ \Gamma_{\phi\to H^\dagger_1 H_2, H^\dagger_2 H_1}\delta_{p1} +\Gamma_{\phi\phi\to H^\dagger_1 H_2,H^\dagger_2 H_1}\delta_{p2} +\Gamma_{\phi\phi\to aa}
\eea
with $\delta_{p1}, \delta_{p2}$ being the Kronecker delta.
 
First,  we get the scattering rate of the inflaton condensate for $\phi\phi\to aa$, as follows, 
\bea
\Gamma_{\phi\phi\to aa}=\frac{1}{8\pi (1+w_\phi)\rho_\phi} \sum_{n=1}^\infty |M^a_n|^2 (E_n\beta^{a}_n),
\eea
with 
\bea
|M^a_n|^2 =  4\lambda^2_\Phi \phi^4_0 |({\cal P}^2)_n|^2,
\eea
and $\beta^{a}_n= \sqrt{1-\frac{m^2_{a}}{E^2_n}}$ and $E_n=n \omega$. For $\phi\gg f_a$ during reheating, the effective mass of the axion  is given by $m_{a}=\sqrt{\lambda_\Phi} \phi$. Since $\omega=0.847\sqrt{\lambda_\Phi} \phi_0$, so $2\omega> m_a$, $\phi\phi\to aa$ is open kinematically.
After the Fourier expansion of the inflaton condensate, the averaged scattering rate for $\phi\phi\to aa$ is given by
\bea
\langle \Gamma_{\phi\phi\to aa}\rangle =\frac{9\lambda^2_\Phi \phi^2_0 \omega}{2\pi m^2_\phi}\, \Sigma_a \bigg\langle \bigg(1-\frac{m^2_a}{\omega^2n^2}\bigg)^{1/2}\bigg\rangle \label{scatteringrate1}
\eea
with $\Sigma_a=\sum^\infty_{n=1}n |({\cal P}^2)_n|^2$. 

Similarly, from the Higgs-portal interactions of the PQ inflaton, ${\cal L}_{\rm int}=-\lambda_{i\Phi}|H_i|^2|\Phi|^2$, we obtain the scattering rate of the inflaton condensate, as follows,
\bea
\Gamma_{\phi\phi\to H^\dagger_iH_i}=\frac{1}{8\pi (1+w_\phi)\rho_\phi} \sum_{n=1}^\infty |M^{H_i}_n|^2 (E_n\beta^{H_i}_n), \quad i=1,2,
\eea
with 
\bea
|M^{H_i}_n|^2 = 4\lambda_{H_i\Phi}^2 \phi^4_0|({\cal P}^2)_n|^2
\eea
and $\beta^{H_i}_n = \sqrt{1-\frac{m^2_{H_i}}{E^2_n}}$. Here, the effective masses for the Higgs fields are given by $m^2_{H_i}=m^2_i+\frac{1}{2}\lambda_{H_i\Phi} \phi^2(t),\,  i=1,2$, where $m^2_i$ are the bare Higgs mass parameters.  Thus, after the Fourier expansion, the averaged scattering rate for $\phi\phi\to H^\dagger_iH_i$  is given by
\bea
\langle \Gamma_{\phi\phi\to H^\dagger_i H_i}\rangle =\frac{9\lambda^2_{H_i\Phi}\phi^2_0 \omega}{2\pi m^2_\phi}\, {\hat\Sigma}_H \bigg\langle \bigg(1-\frac{m^2_{H_i}}{\omega^2n^2}\bigg)^{1/2}\bigg\rangle, \quad i=1,2, \label{scatteringrate2}
\eea
with ${\hat\Sigma}_H=\sum^\infty_{n=1}n |({\cal P}^2)_n|^2$. 
We note that we need to take the Higgs-portal coupling to be small enough, namely, $|\lambda_{H_i\Phi}|\lesssim 10^{-5}$, in order to keep the running quartic coupling $\lambda_\Phi$ at the order of $10^{-11}$. 
As $m^2_{H_i}/\omega^2\sim 0.70\lambda^2_{H_i\Phi}/\lambda_\Phi$, $\phi\phi\to H^\dagger_iH_i$ for the lowest inflaton modes is kinematically open if $|\lambda_{H_i\Phi}|\lesssim 1.2\sqrt{\lambda_\Phi}\simeq 4\times 10^{-6}$ for $\lambda_\Phi=1.1\times 10^{-11}$.

For the PQ invariant potential with $p=2$, ${\cal L}_{\rm int}\supset -\kappa_2 H^\dagger_1 H_2 \Phi^2+{\rm h.c.}$, the scattering rates for $\phi\phi\to H^\dagger_1 H_2$ and its hermitian conjugate, are identical, 
\bea
\Gamma_{\phi\phi\to H^\dagger_1H_2}=\frac{1}{8\pi (1+w_\phi)\rho_\phi} \sum_{n=1}^\infty |M^{H}_n|^2 (E_n\beta^{H}_n)=\Gamma_{\phi\phi\to H^\dagger_2H_1},
\eea
with 
\bea
|M^{H}_n|^2 = 4\kappa^2_2 \phi^4_0|({\cal P}^2)_n|^2
\eea
and $\beta^{H}_n = \sqrt{1-\frac{(m_{H_1}+m_{H_2})^2}{4E^2_n}}\sqrt{1-\frac{(m_{H_1}-m_{H_2})^2}{4E^2_n}}$.
Then, after the Fourier expansion, the averaged scattering rate for $\phi\phi\to H^\dagger_1 H_2$  is given by
\bea
\langle \Gamma_{\phi\phi\to H^\dagger_1 H_2}\rangle =\frac{9\kappa^2_2 \phi^2_0 \omega}{2\pi m^2_\phi}\, {\hat\Sigma}_H \bigg\langle \bigg(1-\frac{(m_{H_1}+m_{H_2})^2}{4\omega^2n^2}\bigg)^{1/2}\bigg(1-\frac{(m_{H_1}-m_{H_2})^2}{4\omega^2n^2}\bigg)^{1/2}\bigg\rangle.\label{scatteringrate3}
\eea
We note that the Higgs-portal coupling $\kappa_2$ is similarly bounded to $|\kappa_2|\lesssim 10^{-5}$, in order to keep the running quartic coupling $\lambda_\Phi$ small during inflation.

For the PQ invariant potential with $p=1$, ${\cal L}_{\rm int}\supset -\frac{1}{\sqrt{2}}\kappa_1 H^\dagger_1 H_2 \Phi+{\rm h.c.}$, the decay rates for $\phi\to H^\dagger_1 H_2$ and its hermitian conjugate, are identical, 
\bea
\Gamma_{\phi\to H^\dagger_1 H_2}=\frac{1}{8\pi (1+w_\phi)\rho_\phi} \sum_{n=1}^\infty |M^{H'}_n|^2 (E_n\beta^{H'}_n)=\Gamma_{\phi\to H^\dagger_2 H_1}
\eea
where 
\bea
|M^{H'}_n|^2=\kappa^2_1 \phi^2_0 |{\cal P}_n|^2
\eea
and $\beta^{H'}_n = \sqrt{1-\frac{(m_{H_1}+m_{H_2})^2}{E^2_n}} \sqrt{1-\frac{(m_{H_1}-m_{H_2})^2}{E^2_n}}$. 
After the Fourier expansion,  the averaged decay rate for $\phi\to H^\dagger_1 H_2$  is given by
\bea
\langle\Gamma_{\phi\to H^\dagger_1 H_2}\rangle=\frac{9\kappa^2_1 \omega}{8\pi m^2_\phi}\, \Sigma_{H} \bigg\langle \bigg(1-\frac{(m_{H_1}+m_{H_2})^2}{\omega^2n^2}\bigg)^{1/2}\bigg(1-\frac{(m_{H_1}-m_{H_2})^2}{\omega^2n^2}\bigg)^{1/2}\bigg\rangle, \label{decayrate1}
\eea
with $ {\Sigma}_{H}=\sum_{n=1}^\infty n |{\cal P}_n|^2$.  
Similarly, the $\kappa_1$ term contributes to $\phi\phi\to H^\dagger_1 H_1, H^\dagger_2 H_2$ too, but the corresponding scattering rates are proportional to $\kappa^4_1$, so it turns out to be suppressed by the higher inverse powers of the Planck scale so negligible as compared to the other decay/scattering rates.

The PQ inflaton can be responsible for the generation of masses for the right-handed neutrinos through the Yukawa couplings, which give rise to the decay rate of the inflaton condensate, as follows,
\bea
\Gamma_{\phi\to N_R N_R}=\frac{1}{8\pi (1+w_\phi)\rho_\phi} \sum_{n=1}^\infty |M^N_n|^2 (E_n\beta^N_n)=\Gamma_{\phi\to \overline {N_R}\, \overline {N_R}}
\eea
where 
\bea
|M^N_n|^2= y^2_N \phi^2_0 |{\cal P}_n|^2 E^2_n (\beta^N_n)^2,
\eea
with $\beta^N_n = \sqrt{1-\frac{4m^2_{N}}{E^2_n}}$ and $m_{N}=\frac{1}{\sqrt{2}} y_N \phi$ being the effective masses for the RH neutrinos.  
After averaging over oscillations, we get
\bea
\langle\Gamma_{\phi\to N_R N_R}\rangle= \frac{ 9 y^2_N \omega^3}{8\pi m^2_\phi}\, \Sigma_N \Bigg\langle\bigg(1-\frac{4m^2_{N}}{\omega^2 n^2}\bigg)^{3/2}\Bigg\rangle, \label{decayrate2}
\eea
with $\Sigma_N =\sum_{n=1}^\infty  n^3 |{\cal P}_n|^2$.
We also note that the Yukawa couplings of the RH neutrinos to the PQ field must be chosen to be small enough, namely, $y_N\lesssim 10^{-3}$, in order to make the running effects on the quartic coupling $\lambda_\Phi$ ignorable.
We note that as $4 m^2_{N}/\omega^2\sim 2.79 y^2_N/\lambda_\Phi$, the decay of the lowest inflaton modes into a pair of RH neutrinos are kinematically open if $|y_N|\lesssim 0.60\sqrt{\lambda_\Phi}\sim 2\times 10^{-6}$ for $\lambda_\Phi=1.1\times 10^{-11}$.

Finally, the same Yukawa couplings of the PQ field to the RH neutrinos also give rise to the inflaton scattering, $\phi\phi\to N_R N_R, \overline {N_R}\, \overline {N_R}$, with the corresponding scattering rate,
\bea
\Gamma_{\phi\phi\to N_R N_R}=\frac{1}{8\pi (1+w_\phi)\rho_\phi} \sum_{n=1}^\infty |{M}^{N'}_n|^2 (E_n {\beta}^{N'}_n)= \Gamma_{\phi\phi\to \overline {N_R}\, \overline {N_R}}
\eea
where 
\bea
 |{ M}^{N'}_n|^2=\frac{32 y^4_N m^2_{N} \phi^4_0 ({ \beta}^{N'}_n)^2}{E^2_n}
\eea
and ${\beta}^{N'}_n=\sqrt{1-\frac{m^2_{N}}{E^2_n}}$. Thus, the averaged scattering rate is given by
\bea
\langle \Gamma_{\phi\phi\to N_R N_R}\rangle 
=\frac{36 y^4_N \phi^2_0 }{\pi m^2_\phi \omega}\,{\hat\Sigma}_{N}\bigg\langle  m^2_{N} \bigg(1-\frac{m^2_{N}}{n^2\omega^2}\bigg)^{3/2}\bigg\rangle, \label{scatteringrate4}
\eea
with $ {\hat\Sigma}_N=\sum_{n=1}^\infty n^{-1}|({\cal P}^2)_n|^2$.

\subsection{Reheating temperature}

For $a_{\rm end} \ll a\ll a_{\rm RH}$ where $a_{\rm RH}$ is the scale factor at the time reheating is complete, we can ignore the inflaton decay/scattering rates and approximate the energy density for the inflaton as
\be
\rho_{\rm \phi}(a)\simeq \rho_{\rm end}\left(\frac{a_{\rm end}}{a}\right)^{4}. \label{inflatondensity}
\ee
When the reheating process is dominated by the perturbative decays and scattering processes of the inflaton, we obtain the reheating  temperature during reheating, as follows,
\bea
T_{\rm RH}= \bigg(\frac{30}{\pi^2 g_*(T_{\rm RH})}\bigg)^{1/4} \Big( \frac{4}{3} \sqrt{3}M_P \gamma_\phi\Big).
\eea
where we took the sum of the decay and scattering rates of the inflaton by $\gamma_\phi\equiv \Gamma_\phi/\rho^\frac{1}{4}_\phi$.

We first compute the decay and scattering rates, as follows,
\begin{align}
\gamma_\phi|_{\rm decay} &\simeq \frac{3\sqrt{\pi}}{2}\,  \lambda^{1/4}_\Phi\,\bigg(\frac{\Gamma\big(\frac{3}{4}\big)}{\Gamma\big(\frac{1}{4}\big)}\bigg)^3\,{\rm max}\bigg[0.5y^2_N\Sigma_N {\cal R}^{-1/2}_N, \frac{\kappa_1^2}{\omega^2}\,\Sigma_{H} {\cal R}_H^{-1/2}\bigg], \\
\gamma_\phi|_{\rm scattering} &\simeq  \frac{3}{\sqrt{\pi}\lambda^{3/4}_\Phi}\bigg(\frac{\Gamma\big(\frac{3}{4}\big)}{\Gamma\big(\frac{1}{4}\big)}\bigg) {\rm max}\bigg[\lambda^2_{H_i\Phi}{\hat\Sigma}_H {\hat{\cal R}}_H^{-1/2},  \kappa^2_2 {\hat\Sigma}_H {\bar{\cal R}}_H^{-1/2}, 8y^4_N \frac{m^2_N}{\omega^2}\,{\hat\Sigma}_N {\hat {\cal R}}_N^{-1/2}\bigg].
\end{align}
Here, $\Sigma_N=0.2406$, $\Sigma_{H}=0.2294$, ${\hat\Sigma}_H=0.1255$,  ${\hat\Sigma}_N=0.0310$, and we took the averaged phase space factor for $2m_N, 2m_{H_i}\gg w$ by ${\cal R}_N\equiv 4m^2_N/w^2$, ${\hat {\cal R}}_N\equiv m^2_N/w^2$, ${\cal R}_H\equiv (m_{H_1}+m_{H_2})^2/w^2$, ${\hat {\cal R}}_H\equiv m^2_{H_i}/w^2$ and  ${\bar {\cal R}}_H\equiv (m_{H_1}+m_{H_2})^2/(4w^2)$\cite{Garcia:2020wiy}.
Thus, we can determine the reheating temperature by the inflaton decay/scattering into a pair of Higgs bosons or RH neutrinos, as follows,
{\small
\begin{align}
T_{\rm RH}|_{\rm decay} &\simeq  5.2\times 10^5\,{\rm GeV} \bigg(\frac{100}{g_*(T_{\rm reh})}\bigg)^{1/4}\bigg(\frac{{\rm max}[y_N,\kappa_1/\omega]}{10^{-4}}\bigg)^2\bigg(\frac{\lambda_\Phi}{10^{-11}}\bigg)^{1/4}, \label{decay}  \\
T_{\rm RH}|_{\rm scattering} &\simeq 3.0\times 10^{11}\,{\rm GeV}  \bigg(\frac{100}{g_*(T_{\rm reh})}\bigg)^{1/4} \bigg(\frac{{\rm max}[\lambda_{H_i\Phi},\kappa_2,\sqrt{2}y^2_N m_N/\omega]}{10^{-7}}\bigg)^2 \bigg(\frac{10^{-11}}{\lambda_\Phi}\bigg)^{3/4}\,. \label{scattering}
\end{align}
}
Here, we note that the coefficient $\kappa_1$ of the PQ invariant term, $H^\dagger_1 H_2 \Phi$, is bounded to $|\kappa_1|\lesssim 10^3\,{\rm GeV}(f_a/10^8\,{\rm GeV})$ for electroweak symmetry breaking as discussed in the end of Section 3.2, while $\omega\sim \sqrt{\lambda_\Phi}\phi_0\sim 10^{13}\,{\rm GeV}$, so $\kappa_1/\omega\lesssim 10^{-10}(f_a/10^8\,{\rm GeV})$. Thus, the $\kappa_1$ term is not efficient for reheating as far as $y_N\gtrsim \kappa_1/\omega$.
As a result, we find that a low reheating temperature below $T_{\rm RH}\sim 10^5\,{\rm GeV}$ can be obtained from the inflaton decay with $10^{-10}(f_a/10^8\,{\rm GeV})\lesssim y_N\lesssim 10^{-4}$, but a high reheating temperature  up to $T_{\rm RH}\sim 10^{11}\,{\rm GeV}$ is achievable due to the inflaton scattering with $\kappa_2\lesssim \lambda_{H_i\Phi}\lesssim 10^{-7}$, which is consistent with a small running quartic coupling $\lambda_\Phi$ during inflation \footnote{Here, the condition $\kappa_2\lesssim \lambda_{H_i\Phi}$ comes from the stability along the Higgs fields during inflation, as shown in eq.~(\ref{stability}).}.

\subsection{Axion dark radiation}

The axions produced from the inflaton scattering can remain out of equilibrium.
Then, we get the correction to the effective number of neutrino species, as follows \cite{thermalaxions2,PQpole},
\bea
\Delta N_{\rm eff}= 0.02678\, \bigg(\frac{Y_a}{Y^{\rm eq}_a}\bigg) \bigg(\frac{106.75}{g_{*s}(T_{\rm reh})}\bigg)^{4/3}
\eea
where $Y_a$ is the axion abundance produced from the inflaton scattering, $Y^{\rm eq}_a$ is the axion abundance at equilibrium given by $Y^{\rm eq}_a=\frac{45 \zeta(3)}{2\pi^4 g_{*s}(T_{\rm reh})}$, and $N^{\rm SM}_{\rm eff}=3.0440$ in the SM \cite{Neff}.  Then, we need $Y_a\lesssim 10 Y^{\rm eq}_a$ to be consistent with the current bounds from the Planck satellite, $N_{\rm eff}=2.99\pm 0.17$ \cite{planck2}.

In our model, the ratio of the scattering rates into the axion pair and Higgs pair is given by
\bea
\frac{\Gamma_{\phi\phi\to aa}}{\Gamma_{\phi\phi\to H^\dagger_i H_i}}\simeq \frac{\lambda^2_\Phi}{\lambda^2_{H_i\Phi}}.
\eea 
Thus, for $\lambda_{H_i\Phi}\gtrsim \lambda_\Phi$,  $\phi\phi\to aa$ is subdominant for reheating, but the produced axions can contribute to the effective number of neutrinos, $\Delta N_{\rm eff}$. In this case, the axions produced during reheating make small contributions to $\Delta N_{\rm eff}$ \cite{PQpole}.

On the other hand, when the universe is reheated to a sufficiently high reheating temperature \cite{thermalaxions1,thermalaxions2},
\bea
T_{\rm reh}\gtrsim 1.7\times 10^9\,{\rm GeV}\bigg(\frac{f_a}{10^{11}\,{\rm GeV}}\bigg)^{2.246}\equiv T_{a, {\rm eq}}, \label{axionthermal}
\eea
the axions could become thermalized with the SM plasma.
In this case,  the contribution of the axions to the effective number of neutrino species is given just by the abundance in thermal equilibrium $Y^{\rm eq}_a$, as follows,
\bea
\Delta N_{\rm eff}=\frac{4}{7}\bigg(\frac{T_{a,0}}{T_{\nu,0}}\bigg)^4=\frac{4}{7}\bigg(\frac{11}{4}\bigg)^{4/3} \bigg(\frac{g_{*s}(T_0)}{g_{*s}(T_{a, {\rm eq}})}\bigg)^{4/3} \label{Neff}
\eea
where $T_{\nu,0}, T_{a,0}$ are the neutrino and axion temperatures, respectively, at present, and $g_{*s}(T_0)=3.91$. Thus, we get $\Delta N_{\rm eff}=0.02397$ for $g_{*s}(T_{a, {\rm eq}})=116$ (adding one more Higgs doublet and three right-handed neutrinos to the SM); $\Delta N_{\rm eff}=0.02550$ for $g_{*s}(T_{a, {\rm eq}})=110.75$ (adding one more Higgs doublet to the SM). 
We remark that future CMB experiments such as CMB-S4 \cite{CMBS4} can test the excess in the effective number of neutrinos in the future.

\subsection{PQ symmetry restoration after reheating}

After reheating, the leading order thermal potential for the PQ field is given by
\bea
V_T(\Phi)&=& \frac{1}{24} T^2 \Big(\sum_{b} n_b m^2_{b,{\rm eff}}+\sum_{f} n_f m^2_{f,{\rm eff}} \Big) +\cdots \nonumber \\
&=& \beta T^2|\Phi|^2 +\cdots
\eea
with
\bea
\beta\equiv \frac{1}{24}(4\lambda_{H_1\Phi}+  4\lambda_{H_2\Phi}+6y^2_N). 
\eea
Thus, if $\beta T^2+\mu^2_\Phi>0$, the PQ symmetry would be restored after reheating, so domain-walls and cosmic strings could be formed even after inflation.  Taking $\mu^2_\Phi\simeq -\lambda_\Phi v^2_\Phi$ in eq.~(\ref{PQvev}) and $v_\Phi\simeq f_a$, we get the upper bound on the reheating temperature in order for the PQ symmetry not to be restored after reheating, as follows,
\bea
T_{\rm reh}< \sqrt{\frac{\lambda_\Phi}{\beta}} \, f_a\equiv T_{\rm restore}.
\eea
Therefore, for $\lambda_{H_1\Phi}, \lambda_{H_2\Phi}\sim 10^{-10}$ and $y_N\sim 10^{-6}$, the upper bound on the reheating temperature is given by $T_{\rm restore}\simeq 0.57 f_a$.

\section{Axion kinetic misalignment}

We briefly summarize the evolution of the axion velocity in the post-inflationary period and determine the axion relic density from the axion kinetic misalignment.

\subsection{Post-inflationary evolution of axion velocity}

After the end of inflation, the total Noether charge for the PQ symmetry is conserved approximately, so $a^3 n_\theta=a^3 \phi^2 {\dot\theta}$ is almost constant.  As a result, the PQ Noether charge density from the axion rotation red-shifts at the end of reheating by
\bea
n_\theta(T_{\rm RH})=n_{\theta,{\rm end}}\,\bigg(\frac{a_{\rm end}}{a_{\rm RH}}\bigg)^3
\eea
where $a_{\rm end}, a_{\rm RH}$ are the values of the scale factor at the end of inflation and the reheating completion, respectively.
Then, suppose that the reheating temperature is sufficiently high such that $\phi(a_{\rm RH})>3f_a$, namely, $T_{\rm RH}>T^c_{\rm RH}$, with
\bea
T^c_{\rm RH}&\equiv& \bigg(\frac{90\lambda_\Phi}{8\pi^2 g_*}\bigg)^{1/4} 2f_a \nonumber \\
&=& \bigg(\frac{100}{g_*}\bigg)^{1/4} \bigg(\frac{f_a}{10^{11}\,{\rm GeV}}\bigg)\, (1.2\times 10^8\,{\rm GeV}).
\eea
In this case, using
\bea
\frac{a_{\rm end}}{a_{\rm RH}}=\bigg(\frac{\rho_{\rm RH}}{\rho_{\rm end}}\bigg)^{1/4}, 
\eea
with $\rho_{\rm RH}=\frac{\pi^2}{30} g_*(T_{\rm RH}) T^4_{\rm RH}$ and $\rho_{\rm end}=\frac{3}{2} V_E(\phi_{\rm end})$, we obtain
the PQ Noether charge density at the reheating temperature, as follows,
\bea
n_\theta(T_{\rm RH}) = n_{\theta,{\rm end}}\,\bigg(\frac{\pi^2 g_*(T_{\rm RH})T^4_{\rm RH}}{45 V_E(\phi_{\rm end})}\bigg)^{3/4}. \label{density1}
\eea
Here, $g_*(T_{\rm RH}), g_*(T_*)$ are the number of the effective entropy degrees of freedom at the reheating temperature and the onset of the axion oscillation, respectively. Thus, the  PQ Noether charge density at $T=T_{\rm RH}$ is independent of the reheating temperature, so is the axion abundance. 

When the reheating is delayed such that $T_{\rm RH}<T^c_{\rm RH}$, the energy density of the inflation scales during reheating as 
\bea
\rho_\phi =\rho_{\rm end} \bigg(\frac{a_{\rm end}}{a_c}\bigg)^4 \bigg(\frac{a_c}{a_{\rm RH}}\bigg)^3
\eea
where $a_c$ is the scalar factor when $\phi(a_c)=3f_a$ such that the inflation becomes matter-like for $a>a_c$.  Then, using 
\bea
\frac{a_c}{a_{\rm RH}}=\bigg(\frac{\rho_{\rm RH}}{\rho_{\phi,c}}\bigg)^{1/3}=\bigg(\frac{T_{\rm RH}}{T^c_{\rm RH}}\bigg)^{4/3}, \label{MD}
\eea
we get the PQ Noether charge density at the end of reheating  for $T_{\rm RH}<T^c_{\rm RH}$ as 
\bea
n_\theta(T_{\rm RH})&=&n_{\theta,{\rm end}}\,\bigg(\frac{a_{\rm end}}{a_c}\bigg)^3\bigg(\frac{a_c}{a_{\rm RH}}\bigg)^3 \nonumber \\
&=&n_{\theta,{\rm end}}\ \bigg(\frac{\pi^2 g_*(T_{\rm RH}) T^4_{\rm RH}}{45 V_E(\phi_{\rm end})}\bigg)^{3/4} \bigg(\frac{T_{\rm RH}}{T^c_{\rm RH}}\bigg). \label{density2}
\eea

\begin{figure}[!t]
\begin{center}
 \includegraphics[width=0.45\textwidth,clip]{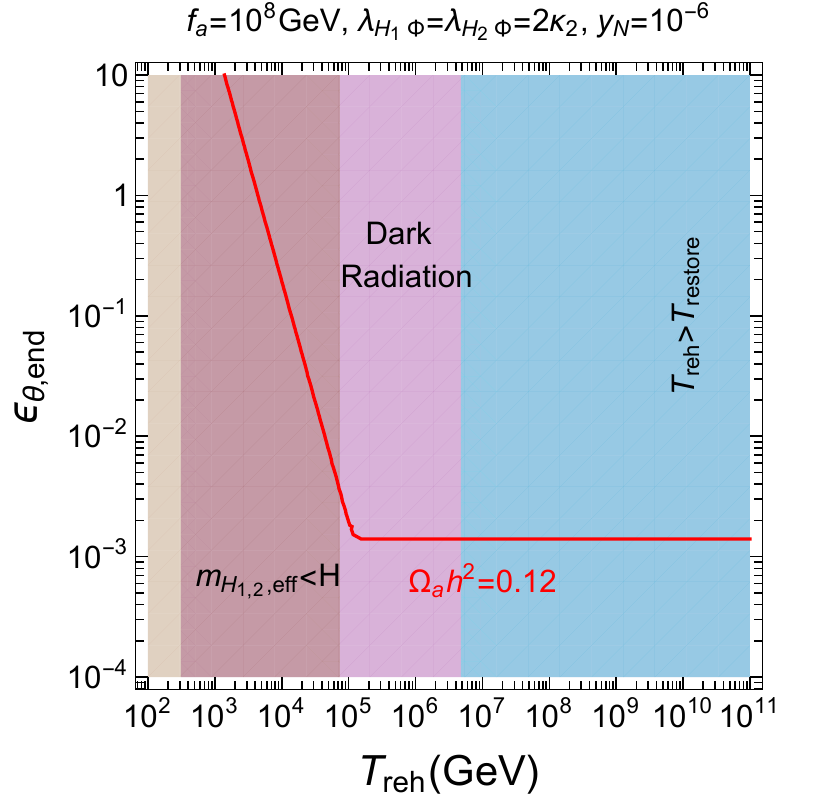}\,\,  \includegraphics[width=0.45\textwidth,clip]{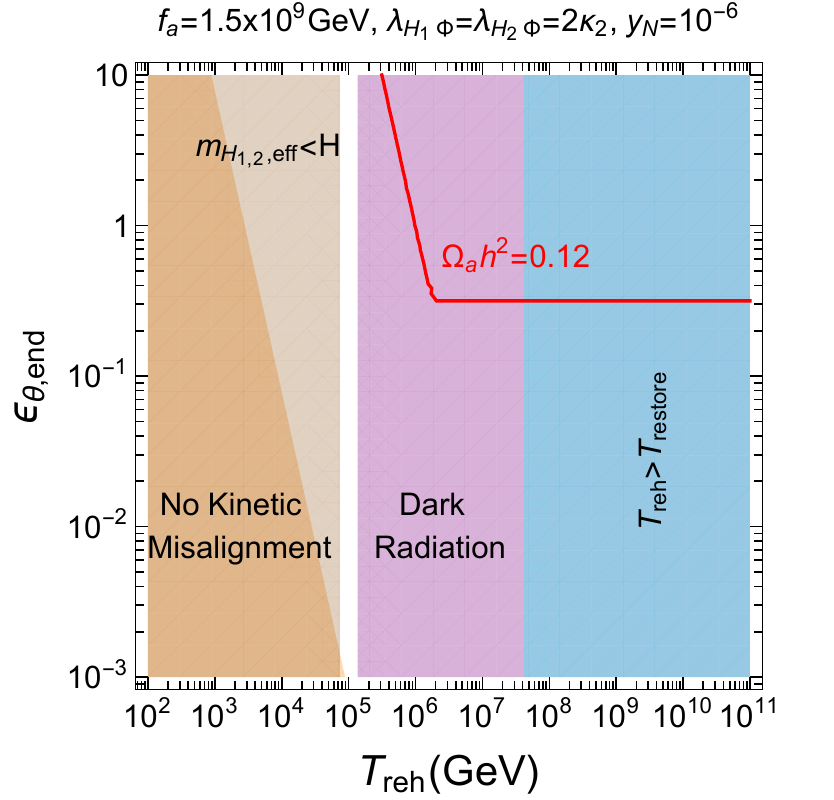}  
 \end{center}
\caption{Reheating temperature $T_{\rm reh}$ vs $\epsilon_{\theta,{\rm end}}$ for axion dark matter with kinetic misalignment. The correct relic density is obtained by the axion kinetic misalignment along the red line. The kinetic misalignment is sub-dominant in orange region, the axions produced from the inflaton scattering were in thermal equilibrium in purple regions, the PQ symmetry is restored after reheating in cyan region, and the effective Higgs masses are smaller than the Hubble scale during inflation in brown region. We took $f_a=10^8\,{\rm GeV}, 1,5\times 10^{9}\,{\rm GeV}$ on the left and right plots, respectively. We chose $\lambda_{H_1\Phi}=\lambda_{H_2\Phi}=2\kappa_2$ and $y_N=10^{-6}$. }
\label{fig:relic1}
\end{figure}

\subsection{Axion relic density}

After the axion gets massive due to the QCD instanton effects and the kinetic energy of the axion is comparable to the potential of the axion, $\frac{1}{2} f^2_a {\dot\theta}^2(T_*)=2m^2_a(T_*) f^2_a $, at the temperature $T=T_*$, the axion is confined to one of the local minima of the axion potential and it starts oscillating for $m_a(T_*) \geq 3 H(T_*)$. Using ${\dot\theta}(T_*)=2m_a(T_*)$, we get the condition for  the axion kinetic misalignment as ${\dot\theta}(T_*)\geq 6H(T_{\rm osc})$, so the onset of oscillation, namely,  $T_*\leq T_{\rm osc}$, where $T_{\rm osc}$ is the temperature of the standard axion oscillation determined by $m_a(T_{\rm osc})=3H(T_{\rm osc})$, with no initial axion kinetic energy.

When the axion kinetic misalignment is a dominant mechanism for determining the axion relic density, we obtain the axion relic abundance by
\bea
\Omega_a h^2 = 0.12 \bigg(\frac{10^9\,{\rm GeV}}{f_a}\bigg) \bigg(\frac{Y_\theta}{40}\bigg)  \label{relic}
\eea
where $Y_\theta$ is the abundance for the axion  given by $Y_\theta=\frac{n_\theta(T_{\rm RH})}{s(T_{\rm RH})}$ with $n_\theta(T_{\rm RH})$  and  $s(T_{\rm RH})$ being the Noether charge density and the entropy density at reheating, respectively. In comparison, the axion abundance, $Y_{a,{\rm mis}}=\frac{n_a}{s}$, determined by the axion misalignment is given by $Y_{a,{\rm mis}}=0.11 (f_a/10^9\,{\rm GeV})^{13/6}$. So, the axion kinetic misalignment is dominant as far as $f_a<1.5\times 10^{11}\,{\rm GeV}$ \cite{Co:2019jts}.

In Fig.~\ref{fig:relic1}, we show the parameter space for the reheating temperature, $T_{\rm reh}$, vs the slow-roll parameter for the axion at the end of inflation, $\epsilon_{\theta,{\rm end}}$, satisfying the correct relic density from the axion in red lines. As compared to the results in Ref.~\cite{PQpole}, we have indicated the region where the PQ symmetry is restored after reheating, namely, $T_{\rm reh}>T_{\rm restore}$, in cyan color, and also showed the region where the effective masses for two Higgs doublets, which correspond to $m_{\pm, {\rm eff}}$ in eq.~(\ref{effmasses}) and $m_{{\tilde A},{\rm eff}}$ in eq.~(\ref{Aeffmass})), are smaller than the Hubble scale during inflation, namely, $m_{H_{1,2},{\rm eff}}<H_I$,  in brown color. In the parameter space where the relic density is explained, the axions produced from reheating can be dark radiation at a detectable level in the future CMB experiments, as shown in the purple region. 
We fixed the axion decay constant to $f_a=10^8, 1.5\times 10^9\,{\rm GeV}$ in the left and right plots, respectively, and chose $\lambda_{H_1\Phi}=\lambda_{H_2\Phi}=2\kappa_2$ and $y_N=10^{-6}$ in common. 

As a result, we find that the low reheating temperature below $10^5\,{\rm GeV}$ is bounded by $m_{H_{1,2},{\rm eff}}>H_I$ due to small Higgs-portal couplings required for reheating, and the high reheating temperature above about $10^7\,{\rm GeV}$ is constrained by the restoration of the PQ symmetry after reheating. If the PQ symmetry is restored after reheating, cosmic strings and domain walls can be formed due to the breaking of the PQ symmetry and  the discrete symmetry of the axion potential, respectively. Those cosmic strings and domain walls are not necessarily problematic if the explicit violation of the PQ symmetry is sizable.

\begin{figure}[!t]
\begin{center}
 \includegraphics[width=0.45\textwidth,clip]{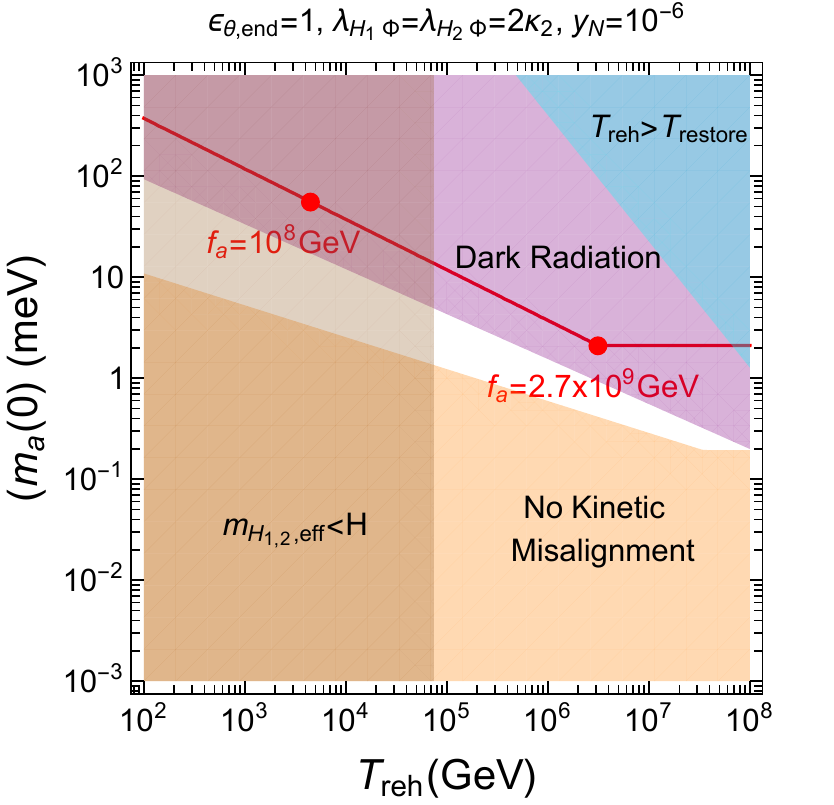}\,\,  \includegraphics[width=0.45\textwidth,clip]{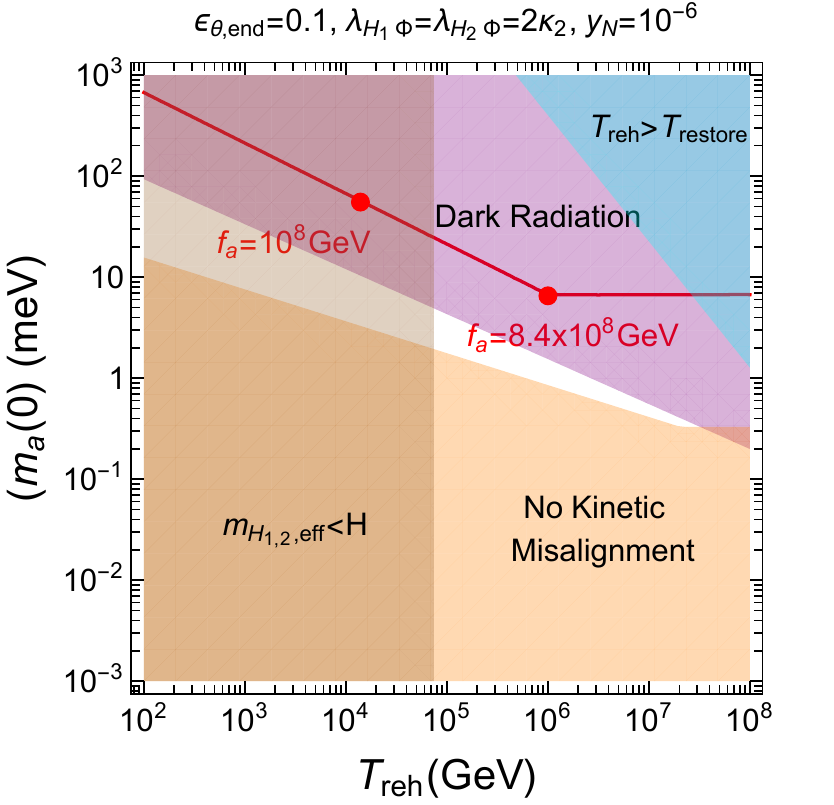}  
 \end{center}
\caption{Reheating temperature $T_{\rm reh}$ vs the axion mass $m_a(0)$  for axion dark matter with kinetic misalignment. The correct relic density is obtained by the axion kinetic misalignment along the red line. The other color codes are the same as in Fig.~\ref{fig:relic1}. We took $\epsilon_{\theta,{\rm end}}=1, 0.1$ on the left and right plots, respectively.  We chose $\lambda_{H_1\Phi}=\lambda_{H_2\Phi}=2\kappa_2$ and $y_N=10^{-6}$. }
\label{fig:relic2}
\end{figure}

In our model, the PQ violating potential gives rise to a nonzero pressure for domain walls formed after QCD phase transition such that they annihilate and never dominate the universe. For a bias term parametrized by $\Delta V=c\,\Lambda^4_{\rm QCD}\times 10^{-10} $ in eq.~(\ref{axionpot}) with $c$ being a dimensionless parameter given by the PQ violating term of order $l$ in the PQ field, as follows, 
\bea
c=|c_{0,l,k}| \bigg(\frac{f_a}{\sqrt{2}q_\Phi M_P}\bigg)^l \bigg(\frac{M_P}{\Lambda_{\rm QCD}}\bigg)^4\times 10^{10},
\eea
the pressure for the domain walls becomes dominant over  the energy density of the domain walls before Big Bang Nucleosynthesis, as far as $\Delta V \gtrsim \frac{\sigma^2}{M^2_P} (t_*/0.1\,{\rm s})$ where $\sigma$ is the tension of domain walls and $t_*$ is the time at which the energy density of the domain walls dominates the radiation energy density, given by $t_*\sim \frac{M^2_P}{\sigma}$ \cite{domainwall}. Thus, for $\sigma\sim \Lambda^2_{\rm QCD} f_a$, there is no domain wall problem as far as $c\gtrsim10^{-3}(f_a/10^9\,{\rm GeV})$, which is satisfied in a consistent parameter space for $l$ and $f_a$ where the condition for the axion quality in eq.~(\ref{axionquality}) and the CMB bound in eq.~(\ref{PQVbound}) are satisfied simultaneously. For instance, it is sufficient to take $l=9(12)$ for $f_a=10^9(10^{12})\,{\rm GeV}$. Otherwise, in the region of the parameter space with $T_{\rm reh}>T_{\rm restore}$ in Fig.~\ref{fig:relic1}, it is relevant to consider the production of axion domain walls and their impact on the axion relic density.

As the Yukawa couplings for the RH neutrinos are increased up to $10^{-4}$, the reheating temperature increases up to $10^5\,{\rm GeV}$ even if  the same Higgs-portal couplings are small and the temperature for restoring the PQ symmetry gets smaller. So, for $y_N>10^{-6}$, the brown and cyan regions are shifted to the right and left in the two plots of Fig.~\ref{fig:relic1}, respectively, resulting in more tight constraints for the axion kinetic misalignment.

In Fig.~\ref{fig:relic2}, we also depict the parameter space for the reheating temperature, $T_{\rm reh}$, vs the axion mass at zero temperature, $m_a(0)$, showing the correct relic density in red lines, for $\epsilon_{\theta,{\rm end}}=1, 0.1$, in the left and right plots, respectively. 
The lower end, $f_a=10^8\,{\rm GeV}$, is imposed by supernova cooling constraint, whereas the upper ends, $f_a=2.7\times 10^9, 8.4\times 10^8\,{\rm GeV}$, in the left and right plots, are shown, because the axion abundance becomes independent of the reheating temperature above those upper ends. We took the same parameters for $\lambda_{H_1\Phi}, \lambda_{H_2\Phi}, \kappa_2, y_N$, as in Fig.~\ref{fig:relic1}.
Thus, we find that there is a parameter space where the axion kinetic misalignment is responsible for the correct relic density while two Higgs fields are  stabilized and the PQ symmetry remains broken after reheating.

Before closing this section, we remark on the isocurvature perturbations of the axion generated during inflation \cite{iso}. The total axion relic density is composed of $Y_a=Y_{a,{\rm kin}}+Y_{a,{\rm mis}}$, with $sY_{a,{\rm kin}}=2m_a n_\theta$ and $sY_{a,{\rm mis}}=\frac{1}{2}m^2_a f^2_a \theta^2_*$. Then, the power spectrum of the isocurvature perturbation at the horizon exit of the mode $k_*$ during inflation depends only on $Y_{a,{\rm mis}}$ is given by
\bea
P_{\rm iso}(k_*)&=&\bigg(\frac{1}{Y_a} \frac{\partial Y_a}{\partial\theta_*}\bigg)^2 \langle\delta \theta_*^2\rangle \nonumber \\
&=& \bigg[\frac{4}{\theta^2_*}\bigg(\frac{Y_{a,{\rm mis}}}{Y_a}\bigg)^2+\frac{1}{4}\bigg(\frac{1}{\epsilon_{\theta,*}}\frac{\partial \epsilon_{\theta,*}}{\partial\theta_*}\bigg)^2\bigg(\frac{Y_{a,{\rm kin}}}{Y_a}\bigg)^2 \bigg] \langle\delta \theta_*^2\rangle
\eea
where the fluctuation of the initial angle is given by
\bea
 \langle\delta \theta_*^2\rangle= \frac{1}{f^2_{a,{\rm eff}}} \,\bigg(\frac{H_I}{2\pi}\bigg)^2.
\eea
with $f_{a,{\rm eff}}\equiv \sqrt{6}\big|\sinh\big(\frac{\phi_*}{\sqrt{6}M_P}\big)\big|$.
Here, we took into account the non-canonical kinetic term for the fluctuation of $\theta$ in eq.~(\ref{Lbkg}) in the Einstein frame, so the effective axion decay constant becomes very large for $\phi_*\gg M_P$ during inflation. The fluctuation of $Y_{a,{\rm kin}}$ with respect to $\theta$ is nonzero, because the initial axion velocity during inflation depends on the initial value of $\theta$ through the explicit PQ breaking term in eq.~(\ref{eom-theta}), namely, the slow-roll parameter $\epsilon_\theta$ for $\theta$. Although the fluctuation of $Y_{a,{\rm kin}}$ also depends on the inflaton field $\phi$ through the PQ charge, $n_\theta$, in eq.~(\ref{PQcharge}), it does not contribute to the isocurvature perturbations because the inflaton fluctuation is adiabatic.  The bound on the isocurvature perturbations from Planck satellite \cite{planck} is given by
\bea
\beta_{\rm iso} \equiv \frac{P_{\rm iso}(k_*)}{P_{\zeta}(k_*)+P_{\rm iso}(k_*)}<0.038,
\eea
at $95\%$ C.L., with $P_{\zeta}(k_*)=A_s=2.1\times 10^{-9}$ at $k_*=0.05\,{\rm Mpc}^{-1}$.
Then, taking $Y_a\simeq Y_{a,{\rm kin}}$ and using
\bea
\bigg(\frac{1}{\epsilon_{\theta,*}}\frac{\partial \epsilon_{\theta,*}}{\partial\theta}\bigg)^2=4(l-2k)^2 \cot^2\Big((l-2k)\theta_*+A_{0,l,k}\Big),
\eea
 we get
\bea
\frac{4}{\theta^2_*} \bigg(\frac{Y_{a,{\rm mis}}}{Y_{a,{\rm kin}}}\bigg)^2+(l-2k)^2 \cot^2\Big((l-2k)\theta_*+A_{0,l,k}\Big)\lesssim 10.
\eea
Here, we assumed a single explicit PQ breaking term at the order of $f^l_a$ in eq.~(\ref{PQpure-PQV}) and used $\epsilon_*\simeq 2\times 10^{-4}$ for $N=60$ in eq.~(\ref{slowroll}), for which $\phi_*\simeq 7.1M_P$ and the effective axion decay constant during inflation becomes  $f_{a,{\rm eff}}\simeq 22M_P$.
Then, the above result leads to $Y_{a,{\rm mis}}/Y_{a,{\rm kin}}\lesssim 5$ and $|(l-2k) \cot((l-2k)\pi+A_{0,l,k})|\lesssim 3$ for $\theta_*=\pi$, which are consistent with the dominance of the axion kinetic misalignment, as far as $|l-2k|\lesssim 3$ is taken or a mild fine-tuning between $\theta_*$  and the phase shift $A_{0,l,k}$ is made for a small value of $|\cot((l-2k)\theta_*+A_{0,l,k})|$ with a large $|l-2k|$. As a result, in our model, the isocurvature perturbations are sufficiently small due to a large value of the effective axion decay constant during inflation.

\section{Conclusions}

We presented a consistent framework with the $U(1)$ PQ symmetry for setting the initial axion velocity at the end of inflation. Including a PQ complex scalar field and an extra Higgs doublet conformally coupled to gravity, we obtained the PQ anomalies from the SM quarks to solve the strong CP problem by the axion and we achieved a slow-roll inflation dominantly by the PQ invariant potential. We also added three RH neutrinos for seesaw mechanism for neutrino masses. Assuming an explicit violation of the PQ symmetry at the Planck scale in the scalar potential due to quantum gravity effects, we showed that a sufficiently large initial axion velocity can be obtained at the end inflation while the axion quality problem is absent. 

Focusing on the pure PQ inflation where the radial mode of the PQ field is responsible for inflation near the pole of its kinetic term, we obtained successful inflationary predictions for a small running quartic coupling for the PQ field and got a sufficiently high reheating temperature determined dominantly by the Higgs-portal couplings to the PQ field.
We also showed that there is a consistent parameter space for the post-inflation era where the axion kinetic misalignment is dominant for the axion relic density. We found that the reheating temperature can be constrained by the interplay of the stability conditions for the Higgs fields during inflation and the non-restoration of the PQ symmetry after reheating. Namely, the former requires the Higgs-portal couplings or the reheating temperature to be sufficiently large while the latter favors small Higgs-portal couplings or reheating temperature.

\section*{Acknowledgments}

We are supported in part by Basic Science Research Program through the National
Research Foundation of Korea (NRF) funded by the Ministry of Education, Science and
Technology (NRF-2022R1A2C2003567). This research was supported by the Chung-Ang University research grant in 2019.
AM acknowledges support by the DFG Emmy Noether Grant No. PI 1933/1-1.




\end{document}